\documentclass[aps,amssymb,amsmath,twocolumn,pra,superscriptaddress,floatfix]{revtex4-1}
\usepackage{amsmath}
\usepackage{latexsym}
\usepackage{amssymb}
\usepackage{bm}
\usepackage{graphics, epstopdf}
\usepackage{color}
\usepackage{newlfont}
\usepackage{amsfonts}
\usepackage{amsthm}
\usepackage{graphicx}
\usepackage{epsfig}

\newcommand*{\Comb}[2]{{}^{#1}C_{#2}}%

\newcommand{\ket}[1]{|{#1}\rangle}
\newcommand{\bra}[1]{\langle{#1}|}

\usepackage[up]{subfigure}

\newcommand{\be}{\begin{equation}}
\newcommand{\ee}{\end{equation}}
\newcommand{\bc}{\begin{center}}
\newcommand{\ec}{\end{center}}
\newcommand{\bea}{\begin{eqnarray}}
\newcommand{\eea}{\end{eqnarray}}
\newcommand{\ba}{\begin{array}}
\newcommand{\ea}{\end{array}}

\usepackage{times}

\begin{document}
\title{Enhancing robustness of multiparty quantum correlations using weak measurement}
%
\author{Uttam Singh}
\email{uttamsingh@hri.res.in}
\author{Utkarsh Mishra}
\email{utkarsh@hri.res.in}
\affiliation{Quantum Information and Computation Group,\\
Harish-Chandra Research Institute, Chhatnag Road, Jhunsi, 
Allahabad 211 019, India}
\author{Himadri Shekhar Dhar}
\email{dhar.himadri@gmail.com}
\affiliation{School of Physical Sciences, Jawaharlal Nehru University, New Delhi 110067, India}

\date{\today}

\begin{abstract}

Multipartite quantum correlations are important resources for the development of quantum information and computation protocols.
However, the resourcefulness of multipartite quantum correlations in practical settings is limited by its
fragility under decoherence due to environmental interactions. Though there exist protocols to protect
bipartite entanglement under decoherence, the implementation of such protocols for multipartite quantum correlations
has not been sufficiently explored. Here, we study the effect of local amplitude damping channel on the
generalized Greenberger-Horne-Zeilinger state, and use a protocol of optimal reversal quantum weak
measurement to protect the multipartite quantum correlations. We observe that the weak measurement reversal protocol enhances
the robustness of multipartite quantum correlations. Further it increases the critical damping value that corresponds to 
entanglement sudden death. To emphasize the efficacy of the technique in protection of multipartite quantum correlation,
we investigate two proximately related quantum communication tasks, namely, quantum teleportation in a one sender,
many receivers setting and multiparty quantum information splitting, through a local amplitude damping channel.
We observe an increase in the average fidelity of both the quantum communication tasks under the weak measurement reversal protocol.
The method may prove beneficial, for combating
external interactions, in other quantum information tasks using multipartite resources.

\end{abstract}

\maketitle
\section{Introduction}

Quantum correlation is an intrinsic aspect of quantum theory that enables the manifestation of several interesting phenomena
beyond the realms of the classical world. The quintessential form of quantum correlation is entanglement which is an important
resource in various quantum information and computational protocols \cite{HRMP}.
The fundamental need for scalability of quantum information processing and computation protocols require
the generation of controlled
quantum correlations distributed over large number of subsystems \cite{multi}. 
In particular, multipartite quantum correlation is an indispensable resource
in one-way quantum computing \cite{briegel}, secret-sharing protocols \cite{ss, hill} and quantum communication \cite{asdphysnews}.
Recent experimental developments in quantum mechanics have enabled the generation of small multipartite entangled states 
to simulate one-way quantum computers \cite{cluster}, graph states \cite{graph} and open-destination quantum teleportation \cite{tele},
using trapped ions \cite{ion} and photons \cite{cluster, graph, tele, photons}.  

The practical realization of quantum information and computation protocols using multiparty quantum systems
is severely challenged due to decoherence caused by the interaction of the system with the environment. Such interactions create
superfluous quantum correlations between the system and the environment leading to information being scattered in the intractable
Hilbert space of the environment.
Therefore, the fragility of multipartite quantum correlations makes the generation and preservation of
quantum correlations in any quantum system a daunting task for experimentalists. From a theoretical point of view, the lack
of a unique characterization of multipartite quantum correlations in quantum systems obstructs definitive study of
decoherence-induced loss of correlations \cite{mintert}.  
The decoherence models used to study the robustness of multipartite states, are mostly local; i.e., each
subsystem of the state interacts independently with the environment \cite{mintert, dur, somsubro}. 
Therefore, in order to successfully implement a quantum protocol, the degradation of the quantum correlations, through
local decoherence channels, must be suppressed during the application of the protocol.

Various schemes such as distillation protocols \cite{BBPSW} and quantum error correction \cite{PS} have
traditionally been used to protect entanglement under local decoherence (cf. \cite{more}). The efficiency of these methods depend
on the robustness of the entanglement in the initial state.
A more recent approach to tackling local decoherence is using the quantum zeno effect \cite{zeno} and the reversibility of 
quantum weak measurement \cite{qubit, weak1}. The experimental viability of implementing quantum weak measurements
makes the latter method an elegant approach to counter decoherence. The suppression of decoherence on a single qubit using
the reversal of weak measurement has been experimentally exhibited \cite{weak2}.
Further, the protection of bipartite entanglement in two-qubit systems using a reversal weak measurement scheme has been
studied \cite{sun} and experimentally demonstrated \cite{kim}. The method has also
been extended to include two-qutrit quantum correlations \cite{li} and few-body quantum communication tasks \cite{ind}.
A natural progression of the weak measurement approach is to consider the
suppression of the degradation of multipartite quantum correlations under local decoherence as they are desirable for 
applications in scalable quantum information and computation protocols.
An obvious difficulty in designing a weak reversal multipartite decoherence-protection scheme is the simultaneous characterization of the
quantum correlation measure and the weak measurement technique in the multipartite setting.

We use the local amplitude damping channel (LADC) as our decoherence model, and study its effect on the generalized
Greenberger-Horne-Zeilinger (gGHZ) state \cite{GHZ}. Such a channel produces a mixed state of the
system from an initial multipartite pure entangled state, rendering procedural characterization of pure state multipartite
quantum correlations irrelevant. Hence, we characterize the mixed state multipartite quantum correlations using a
multipartite extension of logarithmic negativity \cite{ln} and a measure of global entanglement called the Mayer-Wallach
measure \cite{mw}.
We then apply a multipartite generalization of the decoherence-protection protocol based on the reversibility of weak
measurement. We observe that the protocol makes the multipartite quantum correlations robust against LADC.
This is also evident from the enhanced critical damping value that corresponds to entanglement sudden death.
To further elucidate the efficiency of the protocol, we investigate quantum teleportation \cite{teleorig} in a one sender,
many receivers setting and multiparty quantum information splitting \cite{hill,split} using the amplitude-damped gGHZ state and observe an
increase in the average fidelity of teleportation and information splitting under the weak measurement reversal protocol.

The paper is organized in the following way. In Sec.\,\ref{wk}, we discuss the weak measurement reversal protocol to be
applied to the initial multipartite state under consideration. The characterization of the multipartite quantum correlation
measure is done in Sec.\,\ref{meas}. The results for the suppression of decoherence and enhancement of multipartite quantum
correlation is shown in Sec.\,\ref{res}. Section \ref{telep} discusses the results of the quantum teleportation and quantum 
information splitting tasks. We
conclude in Sec.\,\ref{con} with a brief discussion of the results and its ramifications.

\section{Weak measurement reversal {protocol}}
\label{wk}

Weak measurements were initially developed for 
pre-selected and post-selected ensembles of quantum systems by Aharonov {\it et al.} \cite{ara} and later generalized to cases
without post-selection \cite{brun}. An alternative definition of a weak measurement is obtained from partial collapse
measurement \cite{qubit, weak1}. 
In such instances, weak measurements  are positive-operator valued measure with limited information access as compared
to projective measurement thus allowing the operation to be non-unitarily reversed.
The concept of reversing a partial collapse measurement was initially introduced in \cite{tobecited1} and subsequently
applied to quantum error-correcting codes \cite{weak0}. 
In recent times,  a weak measurement reversal
scheme 
has been developed to suppress the decoherence in a single qubit state \cite{qubit,weak1, weak2},
and subsequently expanded to protect quantum correlations in two qubit states \cite{sun, kim, li}.
The underlying principle for the weak measurement reversal scheme is the fact that any partial collapse measurement can
be reversed \cite{weak0, reverse}.

\begin{figure}[htbp]
\centering
\includegraphics[width=80.25 mm]{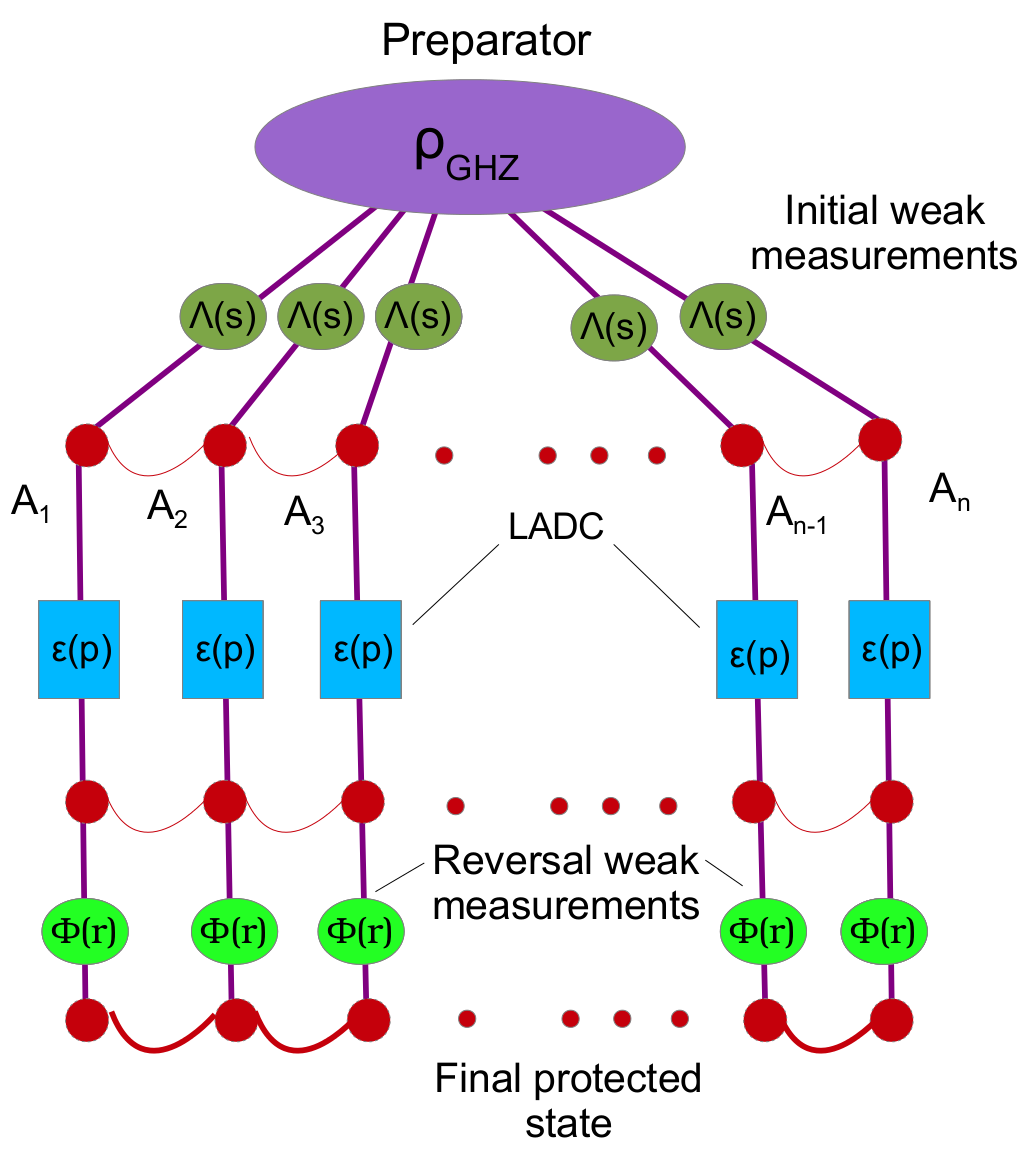}
\caption{Illustration of the weak measurement reversal protocol. The preparator prepares an initial $n$--qubit gGHZ state.
It then applies local weak measurements of strength $s$ on all individual qubits of the $n$--party state. The weak measured
qubits are then individually passed through a LADC with amplitude damping parameter $p$. After local amplitude damping,
each qubit is subjected to a local reversal weak measurement of strength $r$. The multipartite quantum correlation of
the final state is more robust than a $n$--qubit state passing through LADC without weak measurement reversal protocol.  
}
\label{fig:reverse}
\end{figure}

The weak measurement reversal for protection of multipartite quantum correlations can be described by the following protocol.
First, a weak measurement of a given strength, say $s$, is made on the initial multiparty entangled state before the state is subjected to local decoherence.
For an $n$--qubit initial entangled state, the measurement consists of $n$ single--qubit weak measurements, each of strength $s$, acting locally and independently on each qubit.
%
%
Secondly, the post weak measured state undergoes decoherence via a local amplitude damping
channel, with damping parameter $p$, acting independently on each qubit.
Finally, the decohered state is subjected to a reversal weak
measurement, with an optimal reversal strength, $r = r_0$, that is attained by maximizing
the multipartite quantum correlations. 
Again, the reversal weak measurement consists of $n$ single--qubit reversal weak measurements, each of strength $r = r_0$, acting locally and independently on each qubit.
The protocol is graphically illustrated in Fig. (\ref{fig:reverse}).

%
%

In the following parts, we discuss the amplitude damping channel, the weak measurement, and the weak measurement reversal.  

\subsection{Amplitude damping channel}
\label{amp}
Amplitude damping channel is a very useful model of decoherence for studying various quantum phenomena such as
spontaneous emission in quantum optics, energy dissipation in quantum open-systems and capacities of quantum channels.
The decoherence due to amplitude damping results in the coupling of a qubit, locally, to its environment leading to
irreversible transfer of a basis to the other.
The effect of a LADC on a density matrix of a single qubit in computational basis is described by the following map:\\
\begin{eqnarray}
 \label{ampdamp}
 \varepsilon(p):\rho\rightarrow \varepsilon(p)(\rho) &= \left[ \begin{array}{cc}
                 \rho_{00} + p \rho_{11}& \sqrt{1-p}\rho_{01}\\
		 \sqrt{1-p}\rho_{10} & (1-p) \rho_{11}
 \end{array}\right],\\\nonumber
\end{eqnarray}
where $p$ is the amplitude damping parameter and $\rho_{ij}$ ($i, j = 0, 1$) are the elements of $\rho$ in
the computational basis.
Thus, the amplitude damping channel keeps the computational basis state $\ket{0}\bra{0}$ unchanged but transfers the
state from $\ket{1}\bra{1}$ to $\ket{0}\bra{0}$ with probability $p$. The effect of a LADC of strength $p$ on an $n$--qubit state,
say $\rho_n$, is given by the map $\varepsilon(p)^{\otimes n}(\rho_n)$.

\subsection{Weak measurement}
\label{subweak}
The weak measurement, used in our protocol is a positive operator valued measure, consisting of
a set of positive operators $\{M_i(s)\} (i=0,1)$ , such that
\begin{align}
& M_0(s) = \sqrt{s}\ket{1}\bra{1},\nonumber \\
& M_1(s) = \ket{0}\bra{0}+\sqrt{1-s}\ket{1}\bra{1}.
\end{align}
If we discard the outcome of the measurement $M_0(s)$, then the measurement $M_1(s)$ is a null-result weak
measurement of strength $s$, that partially collapses the system to one of the basis states.
The action of the weak measurement operator $M_1(s)$ on a qubit density matrix in computational basis can be
mapped as 
\begin{eqnarray}
 \label{iniweak}
 \varLambda(s):\rho\rightarrow \varLambda(s)(\rho) &= \left[ \begin{array}{cc}
                 \rho_{00} & \sqrt{1-s}\rho_{01}\\
		 \sqrt{1-s}\rho_{10} & (1-s) \rho_{11}
 \end{array}\right],
\end{eqnarray}
where $\rho_{ij}$ ($i, j = 0, 1$) are the elements of $\rho$ in the computational basis.
The quantum map of weak measurement is different from the LADC as it uses post-selection to selectively map
the states. The discarded detections, via some ideal detector, ensures that the operator $M_1(s)$ keeps the computational basis
state $\ket{1}\bra{1}$ unchanged with a probability $(1-s)$ and a norm less than unity. Such post-selection in weak measurements can
increase or decrease quantum correlations in bipartite systems \cite{ent}. The action of the local quantum weak measurement,
of strength $s$, on an $n$--qubit state, say $\rho_n$, is given by the map $\varLambda(s)^{\otimes n}(\rho_n)$.

\subsection{Reversal weak measurement}

The weak measurement
can be reversed using nonunitary operations. For every null-result weak measurement given in Sec. \ref{subweak}, we can obtain a
nonunitary reverse operation $N_0(r)$ with strength $r$, 
that will produce the initial state with some probability. Such an operation is termed as a reversal weak measurement.
Let us consider a measurement using the set of positive operators $\{N_i(r)\} (i=0,1)$, such that
\begin{align}
& N_0(r) = \ket{1}\bra{1} + \sqrt{1-r}\ket{0}\bra{0},\nonumber
\end{align}
\begin{align}
& N_1(r) = \sqrt{r}\ket{1}\bra{1}.
\end{align}
Again if we selectively discard the outcome of measurement $N_1(r)$, then $N_0(r)$ constitutes a reversal weak measurement with
strength $r$. For some optimized reversal weak measurement strength $r=r_0$, corresponding to the original weak
measurement strength $s$,
one can generate the pre-weak measurement state with some probability.
The action of the weak measurement operator $N_0(r)$ on a qubit density matrix in computational basis can be mapped as 

\begin{eqnarray}
 \label{revweak}
 \varPhi(r):\rho\rightarrow \varPhi(r)(\rho) &= \left[ \begin{array}{cc}
                 (1-r)\rho_{00} & \sqrt{1-r}\rho_{01}\\
		 \sqrt{1-r}\rho_{10} & \rho_{11}
		 \end{array}\right],
\end{eqnarray}

where $\rho_{ij}$ ($i, j = 0, 1$) are the elements of $\rho$ in the computational basis.
Similar to the weak measurement, in the reversal measurement, the discarded detections ensures that the operator $N_0(r)$
keeps the computational basis state $\ket{0}\bra{0}$ unchanged with a probability $(1-r)$ and a norm less than unity. 
The action of the local reversal quantum weak measurement
of strength $r$ on an $n$--qubit state, say $\rho_n$, is given by the map $\varPhi(r)^{\otimes n}(\rho_n)$.

\section{multipartite quantum correlations} 
\label{meas}

In our study,
the amplitude damping channel, the weak and the reversal measurements act locally on each subsystem (qubit) of the gGHZ state.
The LADC, in effect, renders the initial pure gGHZ state to a mixed quantum state.
The study of multipartite quantum correlations in this decohered quantum state will require the characterization of
a multipartite measure. Though there are known measures of multipartite entanglement \cite{ggm} which can be
computed in large pure quantum states \cite{our}, these measures do not uniquely generalize to the mixed state regime.
In the following parts we attempt to discuss multipartite quantum correlations that shall prove useful for our study.

\subsection{Multiparty logarithmic negativity}
\label{mln}

Logarithmic negativity (LN) is a useful and computable measure of entanglement introduced by Vidal and Werner \cite{ln} based on
the Peres-Horodecki criteria \cite{peres}. The criteria notes that the negativity of the partial transpose of any bipartite state
is a sufficient condition for entanglement along that particular cut.
For an arbitrary bipartite state, \(\rho_{A:B}\), LN is defined as
\begin{eqnarray}
\label{ln}
E_{\mathrm{LN}}(\rho_{A:B})= \log_2 \left\|\rho_{A:B}^{T_A}\right\|_1
\equiv  \log_2 [ 2 \cal{N}(\rho_{A:B}) + 1 ],
\end{eqnarray}
where
$\cal{N}(\rho_{A:B}) = (1/2)(\left\|\rho_{A:B}^{T_A}\right\|_1- 1)$
is called the ``negativity", and $\left\|\rho_{12}^{T_A}\right\|_1$ is the trace norm of \(\rho_{12}^{T_A}\), which is the partially
transposed state of the bipartite state \(\rho_{A:B}\) with respect to the subsystem $A$.
The negativity, $\cal{N}(\rho_{A:B})$, is thus the sum of the absolute values of the negative eigenvalues of \(\rho_{A:B}^{T_A}\).

For the gGHZ state, the negative partial transpose criteria of entanglement along
all possible bipartite cuts is a sufficient condition. Hence, for gGHZ states, LN is a measure of multipartite entanglement \cite{kempe,suff}. 

\subsection{Global entanglement}

A scalable and functional entanglement monotone for multiqubit pure states, called the global entanglement, was introduced by
Meyer and Wallach \cite{mw}. The Meyer and Wallach measure of global entanglement (MW) is closely related to the per qubit
total nonlocal information \cite{brukner, bren-cai} in the system and hence can be easily related to the loss of nonlocal information
per site under decoherence \cite{afs}. 
The distribution of nonlocal information in a system is known to be related to quantum correlations in a multipartite
quantum system \cite{many}. MW measure is, thus a measure of pure state multipartite quantum correlations.
Interestingly, the connection between the total nonlocal information and tangles can be used to characterize
global entanglement in the decohered gGHZ state.
The MW measure of global entanglement, for an $n$--qubit state in terms of tangles, can be written as \cite{afs}
\begin{eqnarray}
\label{mw}
\mathrm{E}_{gl}^{MW}&=&\frac{1}{n}\Big[2\sum_{i_1<i_2}\tau_{i_1i_2}+3\sum_{i_1<i_2<i_3}\tau_{i_1i_2i_3} + ...\\\nonumber
&+& n\sum_{i_1<...<i_n}\tau_{i_1...i_n}\Big],
\end{eqnarray}
where $\tau_{i_1i_2}$ is the 2--tangle, $\tau_{i_1i_2i_3}$ is the 3--tangle and $\tau_{i_1...i_n}$ is the $n$--tangle.
The term on the right side of Eq. (\ref{mw}), with a multiplicative factor ($n$), is the
expression for total nonlocal information of a quantum system in terms of the tangles. 

The $k$--tangle for an $n$--qubit state can be obtained by generalizing the idea of $3$--tangle \cite{tangle}.
For the states belonging to gGHZ class,
the $n$--tangle ($k = n$) is the only nonzero tangle and, therefore, the sole contributor to Eq. (\ref{mw}) \cite{comments}.
For multiparty states with even $n$, the $n$--tangle
can be expressed in terms of the square of a quantity called the $n$--concurrence \cite{wong},
which is a suitable generalization of concurrence \cite{wooters} for a two qubit state. Hence, for certain 
mixed multipartite states the MW measure of global entanglement can be exactly computed in terms of its $n$--concurrence,
where $n$ is even.
A mathematical disposition of the MW measure for the weak measurement reversal protocol is given in Sec. \ref{res}.

\section{Multipartite quantum correlations under weak measurement reversal protocol} 
\label{res}

In this section, we apply the weak measurement reversal protocol to a local amplitude damped $n$--qubit gGHZ state
and study its effect on the decay of multipartite quantum correlations.
To investigate the effectiveness of the protection protocol, we study the decay of these measures for different values
of amplitude damping parameter $p$ and system size $n$.  
We consider the $n$--qubit generalized GHZ (gGHZ) state \cite{GHZ} as our initial multipartite entangled state, which can be written as
\begin{align}
\label{ghz}
\ket{\psi}_{\mathrm{GHZ}}=\alpha\ket{0}^{\otimes n}+\beta\ket{1}^{\otimes n},
\end{align}
where $\alpha=\cos(\theta/2)$ and $\beta=\sin(\theta/2)$ with $0\leq\theta\leq\pi$.
To apply the local decoherence in the initial state, we pass each of the $n$ qubits of the gGHZ state, separately, through
the amplitude damping channel, given in Sec.\ref{amp}. The action of the amplitude damping channel
on the initial gGHZ state yields the following decohered $n$--qubit state

\begin{align}
\label{gdamp}
& \tilde{\rho}_{\mathrm{GHZ}}=\varepsilon(p)^{\otimes n}(\ket{\psi}\bra{\psi}_{\mathrm{GHZ}})\nonumber\\
 &=|\alpha|^2(\ket{0}\bra{0})^{\otimes n}+\bar{p}^{n/2}[\alpha\beta^*(\ket{0}\bra{1})^{\otimes n}+h.c.]\nonumber\\
 &+|\beta|^2\sum_{k=0}^{n}p^k\bar{p}^{(n-k)}[(\ket{0}\bra{0})^{\otimes k}\otimes(\ket{1}\bra{1})^{\otimes n-k}+ \cal{R}],
\end{align}
where $\bar{p}=1-p$.
$\cal{R}$ denotes all other diagonal terms that can be obtained by permutations of positions of 0 and 1 in the computational basis.

\subsection{Using multiparty logarithmic negativity}
\label{multiln}

We first consider, the calculation of the multipartite logarithmic negativity (Sec.\ref{mln}). To obtain the multipartite LN,
we need to obtain the partial transpose of the state $\tilde{\rho}_{\mathrm{GHZ}}$ across all possible bipartite cuts. 
Let us consider the $m|(n-m)$ cut for partial
transposition. The state after partial transposition along this cut is given by
\begin{align}
&\tilde{\rho}_{\mathrm{GHZ}}^{PT}=\nonumber\\
& |\alpha|^2(\ket{0}\bra{0})^{\otimes n}+\bar{p}^{n/2}
[\alpha\beta^*\ket{\underbrace{0...0}_{m}\underbrace{1...1}_{n-m}}\bra{\underbrace{1...1}_{m}\underbrace{0...0}_{n-m}}+ h.c.]\nonumber\\
 &+|\beta|^2\sum_{k=0}^{n}p^k\bar{p}^{(n-k)}[(\ket{0}\bra{0})^{\otimes k}\otimes(\ket{1}\bra{1})^{\otimes n-k}+ \mathcal{R}],
\end{align}
where the only transformation occurs in the non-diagonal terms of  $\tilde{\rho}_{\mathrm{GHZ}}$.
The partial transposed matrix  $\tilde{\rho}_{\mathrm{GHZ}}^{PT}$ contains a diagonal block of dimension $(2^{n}-2)\times(2^{n}-2)$
that is positive semi-definite. The negative eigenvalue(s) can be obtained from the remaining $2\times 2$ block, which
corresponds to basis vectors $\{\ket{0...0}_{m}\otimes\ket{1...1}_{n-m},\ket{1...1}_{m}\otimes\ket{0...0}_{n-m}\}$. 
The $2\times2$ non-positive semi-definite block can be written as
\begin{eqnarray}
\label{npt}
 B_{\tilde{\rho}_{\mathrm{GHZ}}^{PT}} = \left[ \begin{array}{cc}
                 p^m\bar{p}^{(n-m)}|\beta|^2 & \alpha\beta^*\bar{p}^{n/2}\\
		 \alpha^*\beta\bar{p}^{n/2} & p^{n-m}\bar{p}^m|\beta|^2
 \end{array}\right].
 \end{eqnarray}
The eigenvalue of the above block which contributes to the entanglement, is given by
 \begin{align}
 \label{ev}
  \epsilon_m=\frac{|\beta|^2}{2}\Big[b(p,n,m)-\sqrt{b^2(p,n,m) +4\bar{p}^n\left(\frac{|\alpha|^2}{|\beta|^2}-p^n\right)}\Big],
 \end{align}
where $b(p,n,m)=p^m\bar{p}^{(n-m)}+p^{n-m}\bar{p}^m $.
Using the negative eigenvalue obtained from Eq.(\ref{ev}), we calculate the multipartite LN (defined by Eq.(\ref{ln})).
The condition for nonzero entanglement, i.e., $ \epsilon_m$ to be negative
is given by $p<\min\{1, \left(\frac{|\alpha|}{|\beta|}\right)^{2/n}\}$. Thus, the entanglement of the decohered gGHZ state in the 
bipartition $m|(n-m)$ vanishes for $p \geq p_c$, where $p_c = \min\{1, (|\alpha|/|\beta|)^{2/n} \}$ is the critical damping value.
$p_c$ is found to be independent of the bipartition used to calculate the LN. For $|\alpha|=|\beta|$ case, $p_c = 1$, which
is the case of maximum decoherence. However, numerically the entanglement approaches values
close to zero, before the critical value $p_c = 1$. In our analysis, we
define a new critical value of $p$, denoted by $p_c^a$, as the least value of $p$ above which the entanglement is less than or
equal to $10^{-3}$. Interestingly,
for the case of $|\alpha|<|\beta|$, the multipartite LN becomes zero before the maximal value of decoherence parameter,
$p = 1$, is attained. This phenomenon is called entanglement sudden death (ESD) \cite{esd}.

Now let us consider the operations under the weak measurement reversal protocol, as mentioned in Sec.\ref{wk}. As discussed earlier,
we first apply a weak measurement ($\varLambda(s)^{\otimes n}$), of strength $s$, on the initial $n$--qubit gGHZ state. We then pass the weak measured state through an
amplitude damping channel ($\varepsilon(p)^{\otimes n}$), with damping parameter $p$. Finally, we perform a reversal weak
measurement ($\varPhi(r)^{\otimes n}$), of strength $r$, to obtain the final state. 

The normalized gGHZ state after the application of a weak measurement of strength $s$, is given by
\begin{align}
 \rho_{\mathrm{GHZ}}^{w} &=  \frac { \varLambda(s)^{\otimes n}[\rho_{\mathrm{GHZ}}] }
 {\mathrm{Tr}[\varLambda(s)^{\otimes n}[\rho_{\mathrm{GHZ}}]] } \nonumber\\
 & =|\alpha_1|^2(\ket{0}\bra{0})^{\otimes n} + |\beta_1|^2(\ket{1}\bra{1})^{\otimes n}\nonumber\\
 &~~~~+[\alpha_1\beta_1^*(\ket{0}\bra{1})^{\otimes n} + h.c.],
\end{align}
where $\alpha_1=\alpha/\sqrt{\mathcal{T}_1}$,
$\beta_1=\bar{s}^{n/2}\beta/\sqrt{\mathcal{T}_1}$ and
$\bar{s}=1-s$. $\mathcal{T}_1 = |\alpha|^2+\bar{s}^n|\beta|^2$ is the success probability of weak measurement.
The weak measured state is then passed through a LADC with damping parameter $p$
that acts locally on each qubit. The decohered state is given by
\begin{align}
& \tilde{\rho}_{\mathrm{GHZ}}^{w}
=|\alpha_1|^2(\ket{0}\bra{0})^{\otimes n}+\bar{p}^{n/2}[\alpha_1\beta_1^*(\ket{0}\bra{1})^{\otimes n}+h.c.]\nonumber\\
 &+|\beta_1|^2\sum_{k=0}^{n}p^k\bar{p}^{(n-k)}[(\ket{0}\bra{0})^{\otimes k}\otimes(\ket{1}\bra{1})^{\otimes n-k} + \mathcal{R}_1],
\end{align}
where again, $\cal{R}_1$ denotes all other diagonal terms that can be obtained either by permutations of positions of $0$ and $1$ in
the computational basis. Then we make a reversal weak measurement of strength $r$,
on the state $\tilde{\rho}_{\mathrm{GHZ}}^{w}$. The state after the reversal weak measurement is given by
\begin{align}
\label{proghz}
\tilde{\rho}_{\mathrm{GHZ}}^{wr} &=\frac{1}{\cal{T}_2}\big[|\alpha_1|^2\bar{r}^n(\ket{0}\bra{0})^{\otimes n}
+(\bar{r}\bar{p})^{n/2} [\alpha_1\beta_1^*(\ket{0}\bra{1})^{\otimes n}\nonumber\\
 &+ h.c.] +|\beta_1|^2\sum_{k=0}^{n}(p\bar{r})^k\bar{p}^{(n-k)}[(\ket{0}\bra{0})^{\otimes k}\nonumber\\
 &~~~~~~\otimes(\ket{1}\bra{1})^{\otimes n-k}+ \mathcal{R}_1]\Big], 
\end{align}
where
\begin{align}
\cal{T}_2&=|\alpha_1|^2\bar{r}^n+|\beta_1|^2\sum_{k=0}^{n}\Comb{n}{k}(p\bar{r})^k \bar{p}^{n-k}\nonumber\\
&=|\alpha_1|^2\bar{r}^n+|\beta_1|^2(1-pr)^n,
\end{align}

and $\bar{r}=1-r$. $\mathcal{T}_2$ is the success probability of the reversal weak measurement. The transmissivity or 
the success probability of the final
weak measurement reversal protocol, is given by
\begin{align}
\label{trans1}
\mathcal{T} = \mathcal{T}_1\mathcal{T}_2 =
[ |\alpha|^2\bar{r}^n+|\beta|^2\bar{s}^n(1-pr)^n].
\end{align}

Let us again consider the $m|(n-m)$ bipartition for the partial
transposition of the final post reversal weak  measurement state $\tilde{\rho}_{\mathrm{GHZ}}^{wr}$. 
Following the steps for partial transpose of the non-protected amplitude damped state 
we again obtain a $2\times2$ matrix block that contains the possible non-positive eigenvalue.   
The matrix form of such a block can be found to be
\begin{eqnarray}
 B_{\tilde{\rho}_{\mathrm{GHZ}}^{wr(PT)}} = \frac{1}{\mathcal{T}_2}\left[ \begin{array}{cc}
                 (p\bar{r})^m\bar{p}^{n-m}|\beta_1|^2 & \alpha_1\beta_1^*(\bar{p}\bar{r})^{n/2}\\
		\alpha_1^*\beta_1(\bar{p}\bar{r})^{n/2} & (p\bar{r})^{n-m}\bar{p}^m|\beta_1|^2 
 \end{array}\right].\nonumber\\
 \end{eqnarray}
The eigenvalue, which can be negative is
 \begin{align}
  \epsilon^{wr}_m=&\frac{|\beta_1|^2}{2\mathcal{T}_2}\big[b_1(r,p,m,n)\nonumber\\
  &-\sqrt{b_1^2(r,p,m,n) + 4(\bar{r}\bar{p})^n\left(\frac{|\alpha_1|^2}{|\beta_1|^2}-p^n\right)}\big],
 \end{align}
where $b_1(r,p,m,n)=(p\bar{r})^m\bar{p}^{(n-m)}+(p\bar{r})^{(n-m)}\bar{p}^m$.
The condition for nonzero entanglement, i.e., $ \epsilon^{wr}_m$ to be negative
is given by $p < \min\{1, (\frac{1}{\bar{s}})\left(\frac{|\alpha|}{|\beta|}\right)^{2/n}\}$.
This shows that the weak measurement reversal protocol enhances the critical damping value of $p$
as compared to the unprotected case.
Thus, a higher
damping parameter $p$ is required to remove entanglement.
The optimal value of the reversal weak  measurement strength $r$ is obtained by maximizing the multipartite LN for a fixed value
of the damping strength $p$, the number of qubits $n$, the initial weak  measurement strength $s$ and all possible bipartitions $m$.
We will denote this optimal value of LN by $E_{LN}^{opt}$.
For $|\alpha|<|\beta|$ and specific values of the weak measurement strength $s$, the observed ESD under the amplitude damping
can be suitably circumvented.
 
\begin{figure}[htbp]
\centering
\subfigure[~{\it n} = 4]
{
\includegraphics[width=41.25 mm]{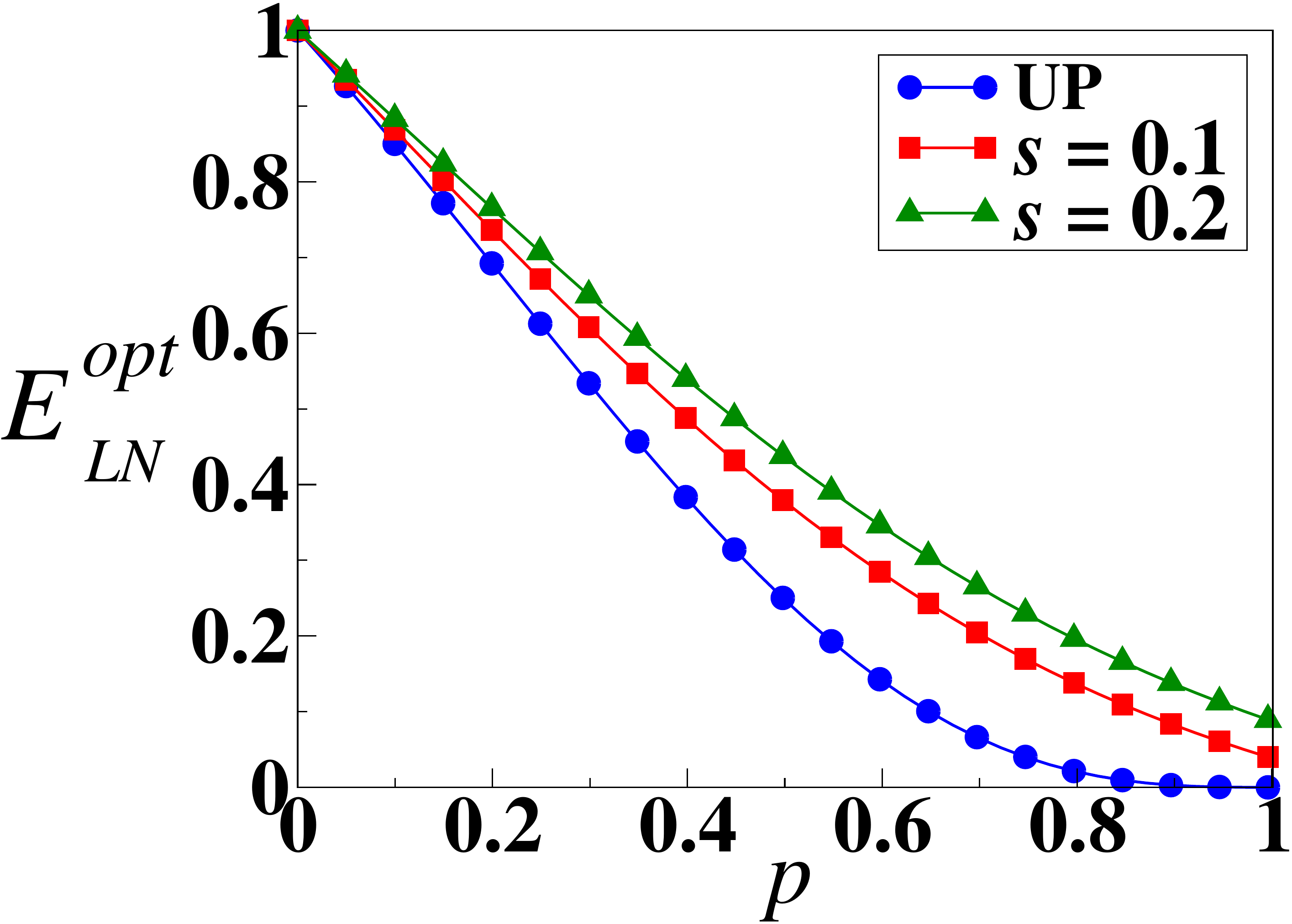}
}
\subfigure[~{\it n} = 8]
{
\includegraphics[width=41.25 mm]{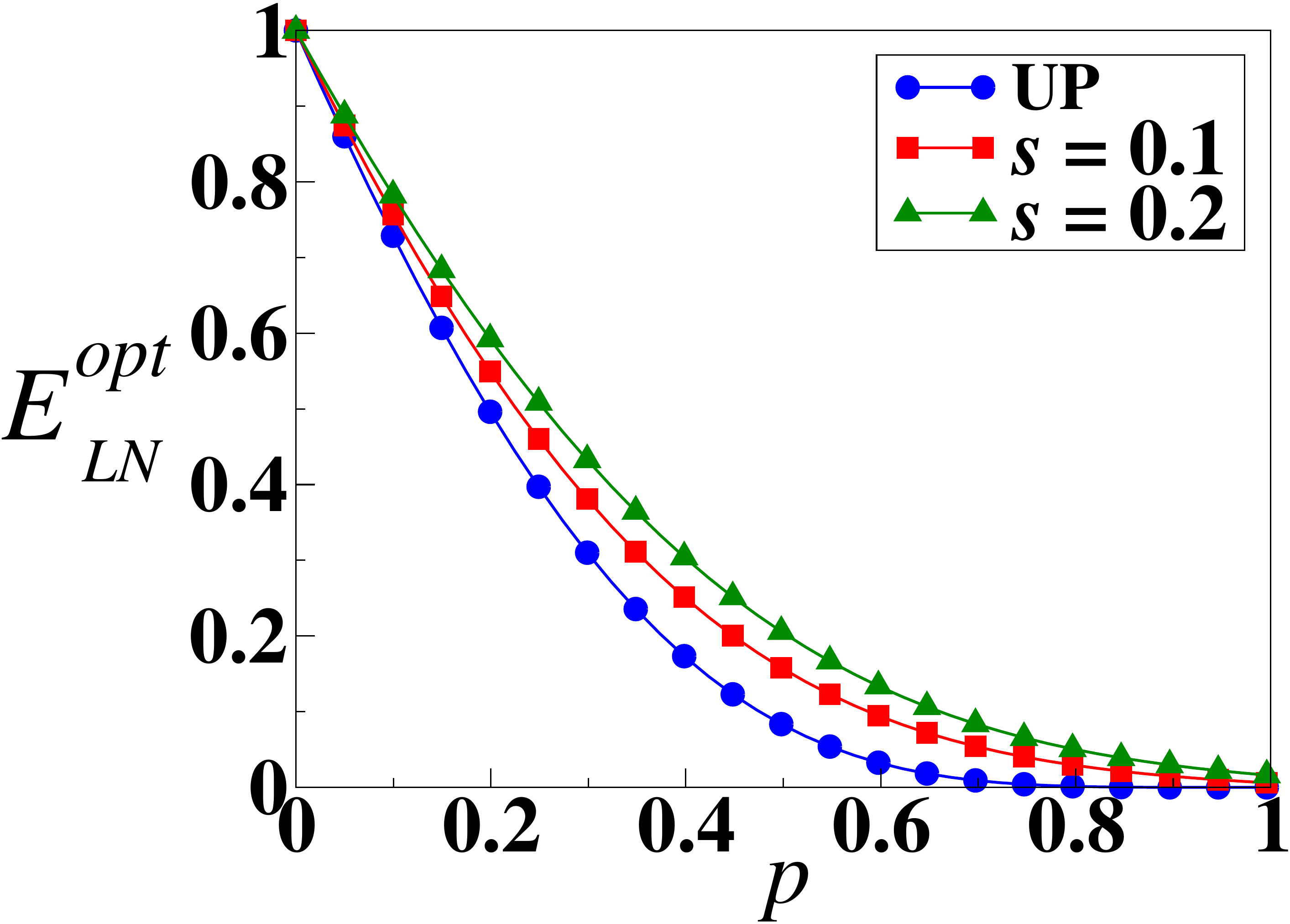}
}
\subfigure[~{\it n} = 12]
{
\includegraphics[width=41.25 mm]{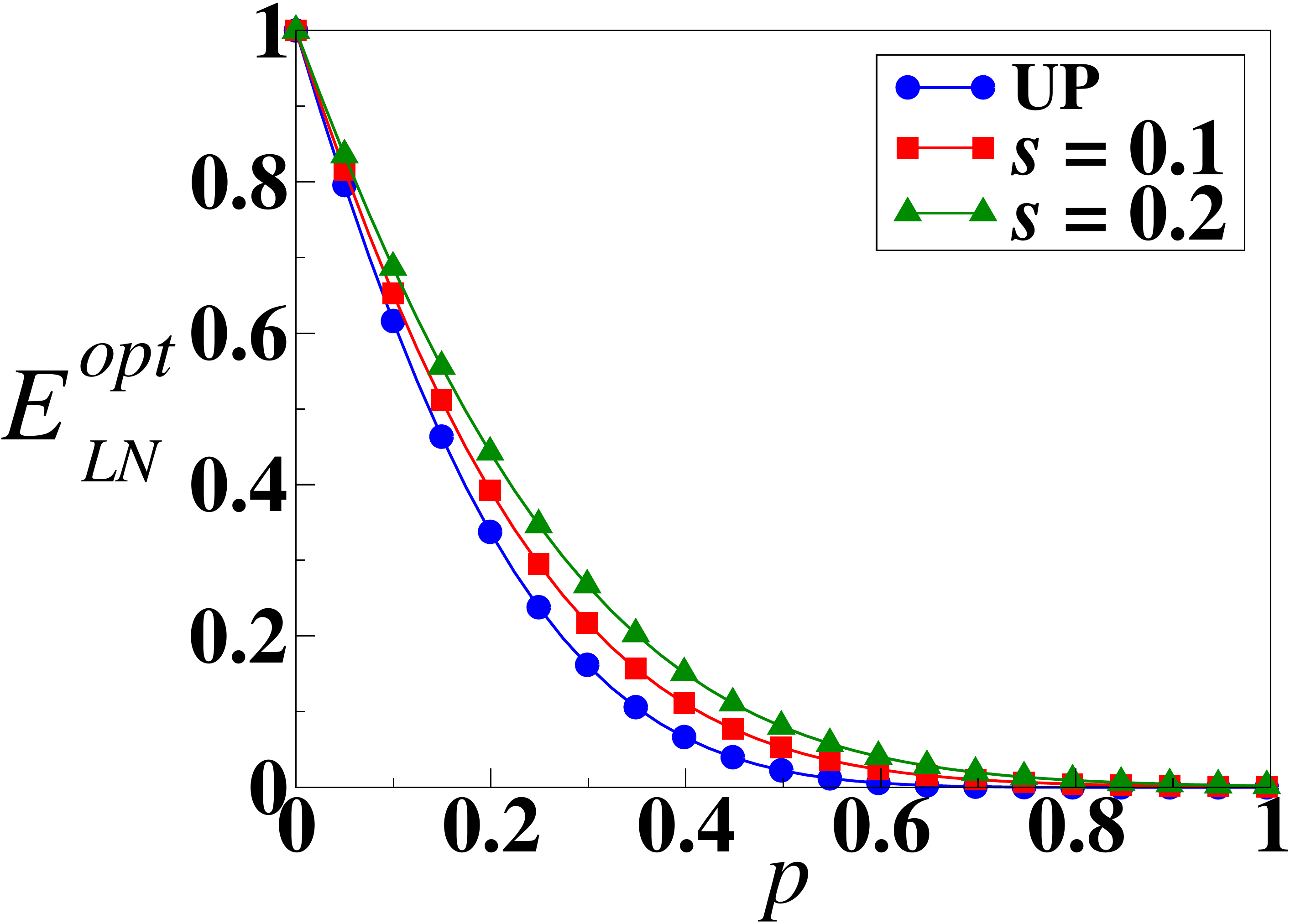}
}
\subfigure[ ~{\it n} = 24]
{
\includegraphics[width=41.25 mm]{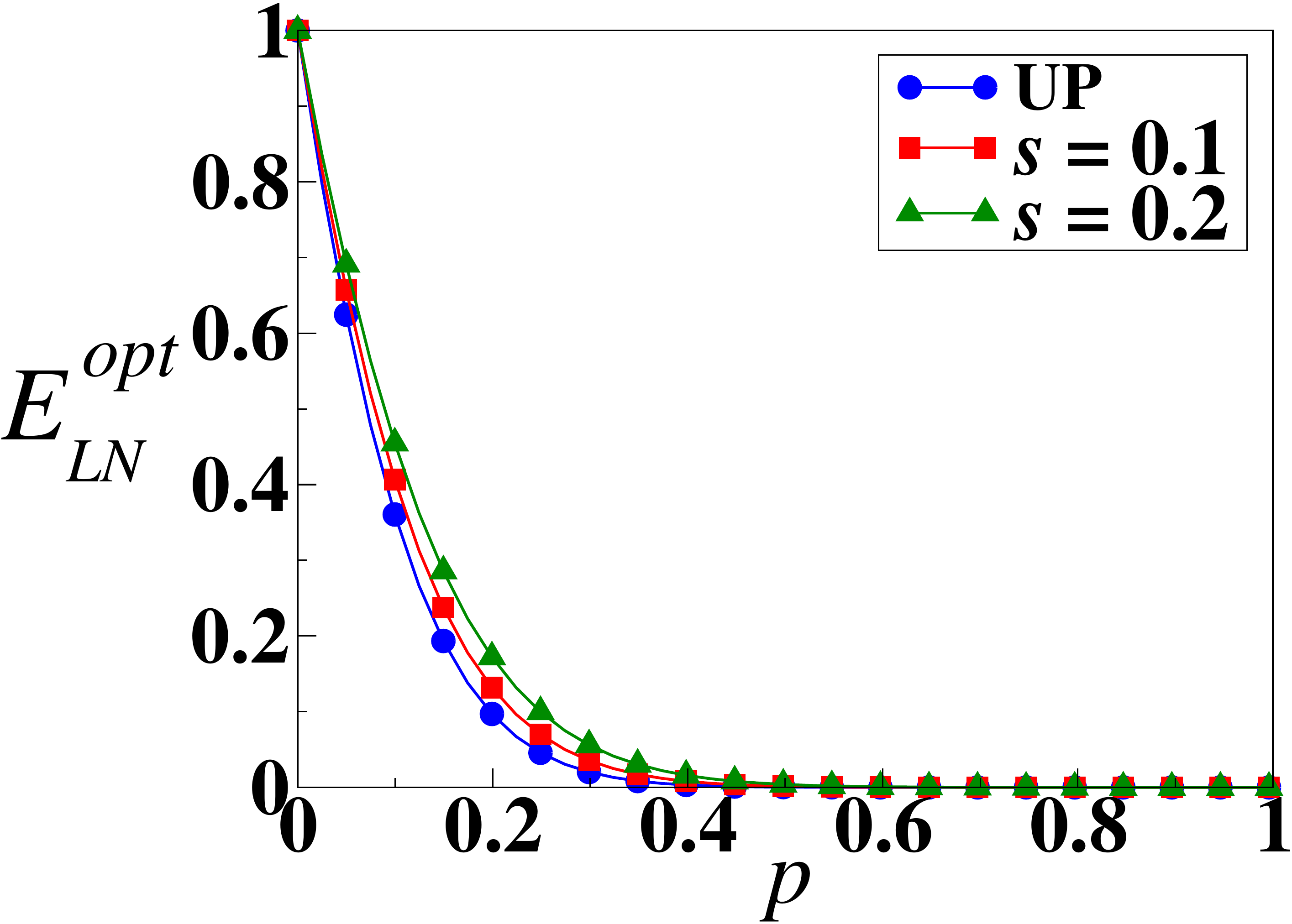}
}
\caption{(Color online) The optimal multipartite logarithmic negativity of the amplitude-damped gGHZ state with $\alpha =\beta = 1/\sqrt{2}$,
under the protection protocol,
as function of amplitude damping parameter $p$, for different values of initial weak measurement strength $s$, system
size $n$ and at the bipartition $m = n/2$. We observe that the decay of $E^{opt}_{LN}$ is retarded and the critical value of $p$ where correlation death occurs
is increased for finite values of $s$. The plot label UP is for the unprotected state ($s$ = $r$ = 0).}
\label{fig1}
\end{figure}

\begin{figure}[htbp]
\centering
\includegraphics[width=50mm]{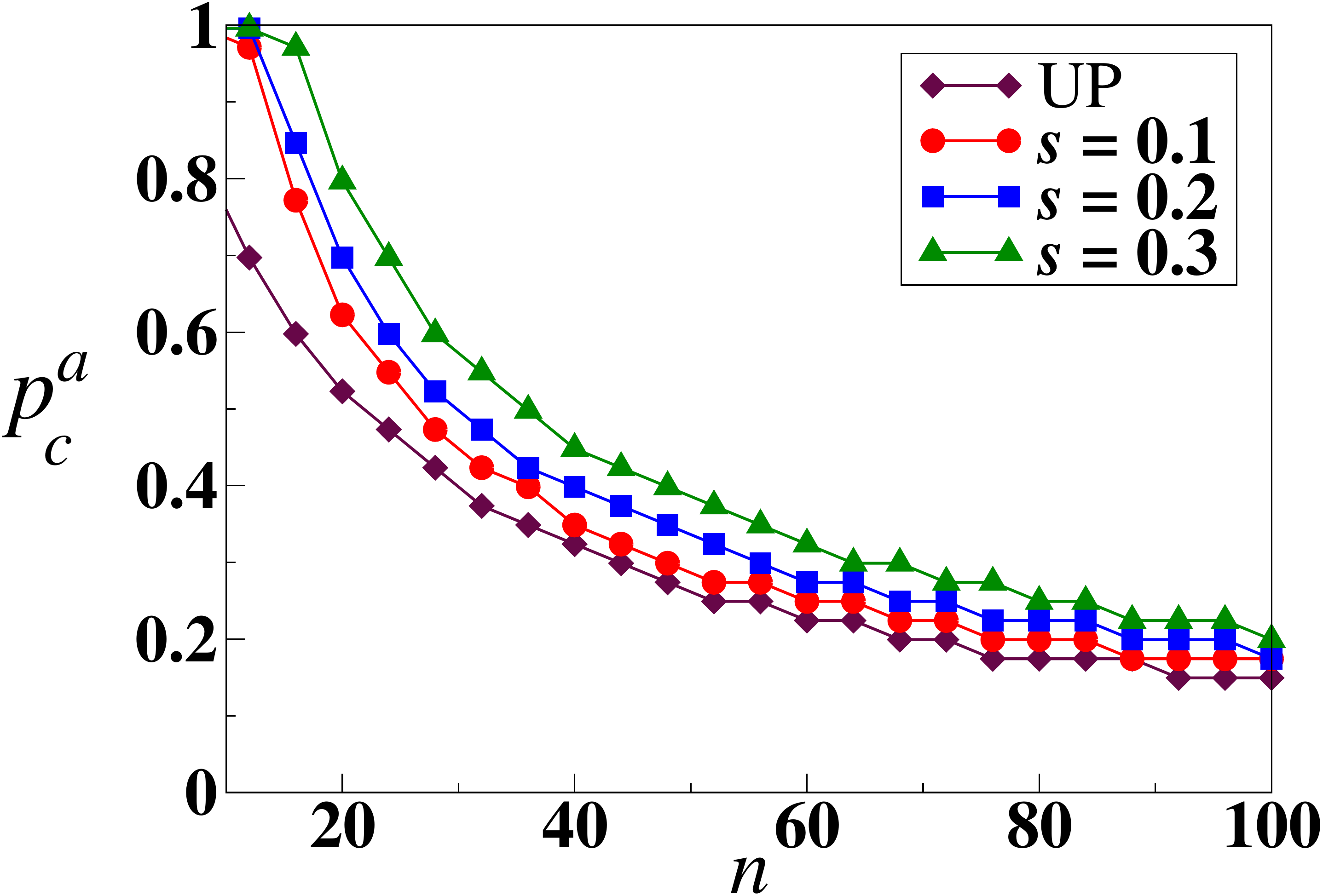}
\caption{(Color online) The critical damping value, $p=p^a_c$, for $E^{opt}_{LN}$ plotted against the number of qubits $n$.
The  plot shows that  $p^a_c$ of the protected state remains above the unprotected state for $n$ = 100, though the difference
diminishes with increasing $n$. The plot label UP is for the unprotected state ($s$ = $r$ = 0).}
\label{cri1}
\end{figure}

Figure (\ref{fig1}) represents the optimal value of the multipartite LN ($E_{LN}^{opt}$)
of the amplitude-damped gGHZ state (with $\alpha =\beta = 1/\sqrt{2}$), under the weak measurement reversal protocol,
as a function of the damping strength $p$, for various
initial weak measurement strength $s$, and for $n$ qubits, at the bipartition $m$ = $n/2$. We observe that the weak measurement
reversal protocol suppresses the decay of multipartite quantum correlations, for different $n$.
From Fig. (\ref{fig1}), we observe that $E_{LN}^{opt}$ decreases with increase of the amplitude damping parameter $p$ and vanishes
after a critical damping value $p^a_c$, that depends on $n$.
%
For smaller values of $n$ ($n=4$ and $n=8$), the weak measurement reversal technique leads to a substantial
increase in the multipartite entanglement for larger values of $s$, making the state more robust against the noise.
However, we observe later that the arbitrary high values of $s\in [0,1]$ are not allowed.
At higher values of $n$, say $n=24$, it is apparent from Fig. \ref{fig1} (also see Fig. \ref{cri1}) that $p^a_c$
becomes smaller and the weak measurement reversal protocol does not sufficiently enhance entanglement (due to limitation on $s$),
rendering the state
less robust against decoherence for very large values of $n$.

Figure \ref{cri1} shows the variation of $p^a_c$ with $n$, for fixed values of the weak measurement strength $s$.
We observe that the values of $p^a_c$ for all $s$, get closer at higher values of $n$, which implies  
that enhancement of the critical value for protected states is lower, rendering the protocol less efficient at very large $n$.

The effective value of the weak measurement strength $s$ also depends
on the success probability of the reversal protocol. To obtain the success probability of the protocol we calculate its transmissivity,
defined in relation (\ref{trans1}). For an initial gGHZ state with $\alpha=\beta=1/\sqrt{2}$, the transmissivity is given by the
expression
$
 \cal{T} = \frac{1}{2}[\bar{r}^n+\bar{s}^n(1-pr)^n].
$
The transmissivity $\cal{T}$ is calculated for the optimal value of the reversal weak measurement strength $r$ for which
$E_{LN}$ is maximal, i.e. $E_{LN} = E^{opt}_{LN}$.
\begin{figure}[htbp]
\centering
\subfigure[~{\it p} = 0.2]
{
\includegraphics[width=41.25 mm]{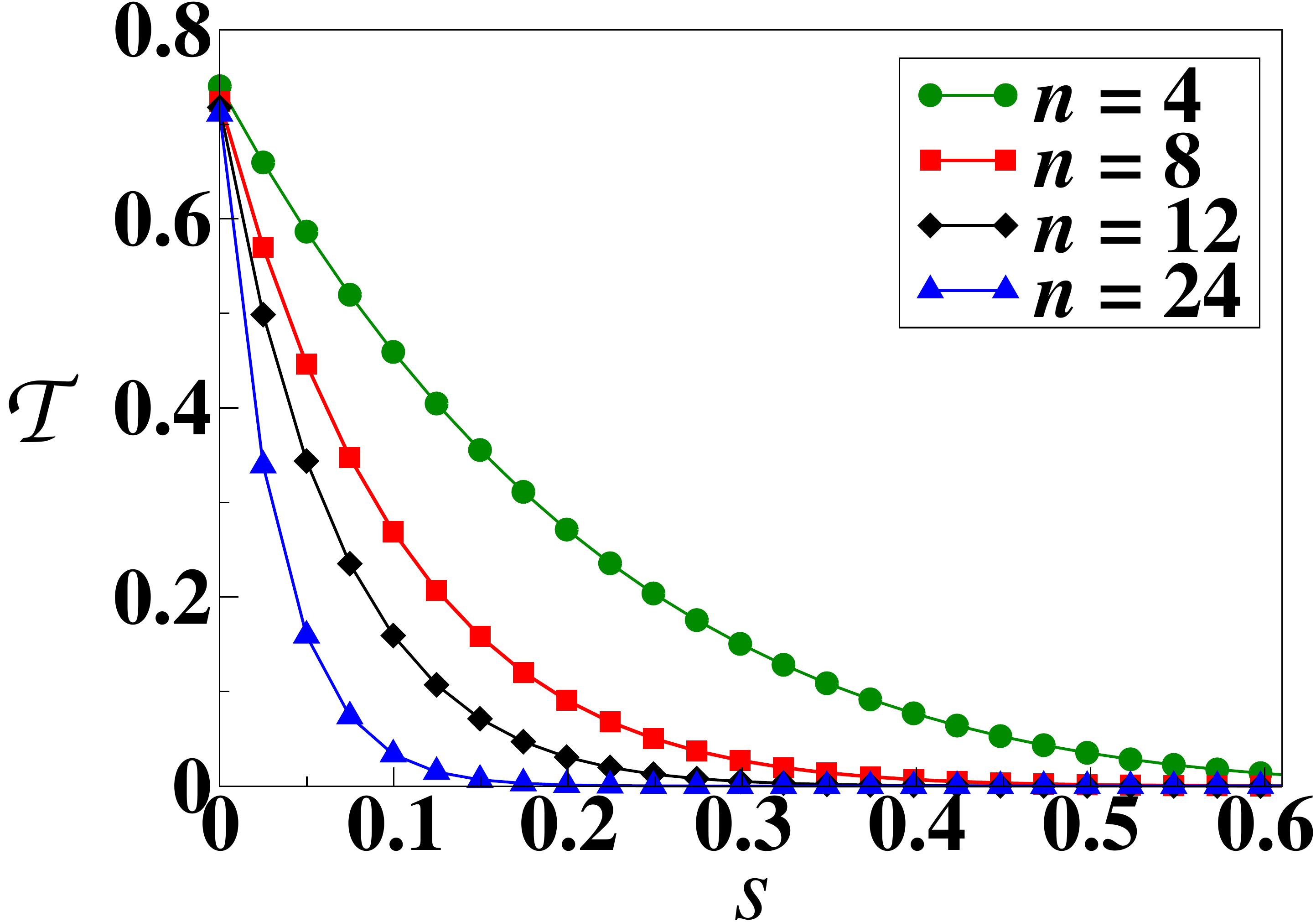}
}
\centering
\subfigure[~{\it p} = 0.3]
{
\includegraphics[width=41.25 mm]{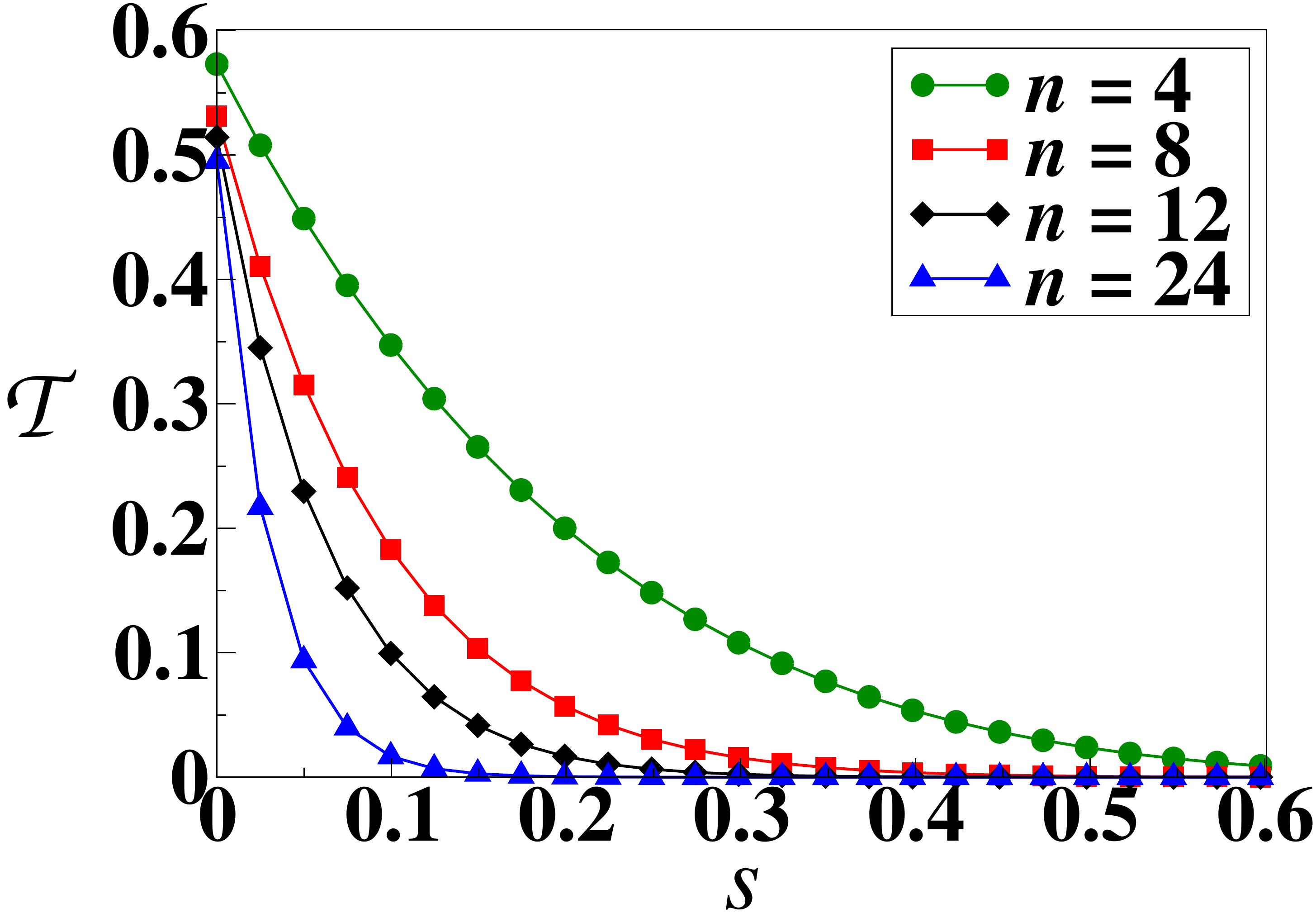}
}
\caption{(Color online) Transmissivity as a function of the initial weak measurement strength $s$ for different values of $n$
and fixed damping $p$. We observe that the success probability reduces for larger number of qubits $n$.}
\label{fig2}
\end{figure}

We observe from Fig. \ref{fig2}, that the success probability reduces with increase in the weak measurement strength $s$ and hence,
the protocol becomes less robust. Thus, for larger values of the initial weak measurement strength, the success probability of
the quantum state reversal using weak measurement becomes lower and hence the number of initial copies of the system needed to
successfully implement the protection protocol increases.

\subsection{Using global entanglement}
\label{gle}

We again consider the $n$--qubit gGHZ state given by Eq. (\ref{ghz}). For the MW measure of global entanglement ($E_{MW}$) in
Eq. (\ref{mw}), it is known that the multipartite quantum correlation for the $n$--qubit gGHZ state is contained only in the
$n$--tangle part of the nonlocal information, since all the other $k$--tangles ($k \neq n$) are zero \cite{comments}. For
an even number of qubits (even $n$), the $n$--tangle
can be calculated using the $n$--concurrence measure $C_n$, as shown in \cite{wong}. 
$C_n$ is a generalization of two qubit measure of concurrence \cite{wooters}. Therefore for an initial, even $n$--qubit gGHZ state
the MW global entanglement is given by the relation
\begin{equation*}
E_{MW}=\sum_{i_1<...<i_n}\tau_{i_1...i_n}=\tau_n = C_n^2,
\end{equation*}
where $C_n$ is defined as
\begin{align}
 \label{conc}
 C_n=\max\Big[0,\{\sqrt{\lambda_1}-\sum_{i=2}^{2^N}\sqrt{\lambda_i}\}\Big],
\end{align}
where $\lambda_i$'s are the eigenvalues of the matrix $R_n=\rho {\sigma_y}^{\otimes n}\rho^*{\sigma_y}^{\otimes n}$
(for any density matrix $\rho$) in decreasing order, 
and $\sigma_y$ is the Pauli matrix. The gGHZ state, after passing through local amplitude
damping channel is given by Eq. (\ref{gdamp}).
The eigenvalues of $R_n=\tilde{\rho}_{\mathrm{GHZ}}(\tilde{\rho}^{sc}_{\mathrm{GHZ}})$ are give by
\begin{align}
 &\lambda_1=|\beta|^2\bar{p}^n\Big[(2|\alpha|^2+|\beta|^2p^n)+\sqrt{4|\alpha|^2(|\alpha|^2+|\beta|^2p^n)}\Big],\nonumber\\
 &\lambda_2=|\beta|^2\bar{p}^n\Big[(2|\alpha|^2+|\beta|^2p^n)-\sqrt{4|\alpha|^2(|\alpha|^2+|\beta|^2p^n)}\Big],\nonumber\\
 &\lambda_j=|\beta|^4(p\bar{p})^n, \{j=3,4,..,2^n\},
\end{align}
where $\tilde{\rho}^{sc}_{\mathrm{GHZ}}$ is the spin conjugated state obtained after applying
${\sigma_y}^{\otimes n}$ on $\tilde{\rho}^*_{\mathrm{GHZ}}$.
Therefore, the $n$--concurrence is given by
\begin{align}
 &C_n(\tilde{\rho}_{\mathrm{GHZ}})=\max\Big[0,\big\{\sqrt{\lambda_1}-\sqrt{\lambda_2}-\sum_{j=3}^{2^n}\sqrt{\lambda_j}\big\}\Big]\nonumber\\
 &=\max\Big[0,\big\{2|\alpha||\beta|\bar{p}^{n/2}\big[1-(2^{n-1}-1)\frac{|\beta|}{|\alpha|}p^{n/2}\big]\Big\}\Big].
\end{align}

The above expression shows that the $n$--concurrence is zero for $p\geq p^{MW}_c$, where $p^{MW}_c = \min\{1, \left(\frac{|\alpha|}{
|\beta|(2^{n-1}-1)}\right)^{2/n} \}$, and positive definite otherwise. Hence ESD occurs at the critical damping parameter
value, $p = p^{MW}_c$. In the limit of very large $n$, the critical value 
$p^{MW}_c$ becomes $1/4$, independent of the parameters of the initial gGHZ state. The $n$--concurrence indicates that the gGHZ
can sustain decoherence upto $p$ = $0.25$ for very large $n$, unlike the logarithmic negativity which, in principle
can sustain decoherence upto $p$ = $1$.
The global entanglement, $E_{MW}$ of the state $\tilde{\rho}_{\mathrm{GHZ}}$ is given by
$C_n^2(\tilde{\rho}_{\mathrm{GHZ}})$. 
We define the numerical critical value of $p$ in the case of $E_{MW}$, $p^{MWa}_c$, as the value of $p$ below
which the $E_{MW}$ is less than $10^{-3}$.

Now the state after the weak measurement reversal protocol is given by (\ref{proghz}).
The eigenvalues of $R'_n=\tilde{\rho}_{\mathrm{GHZ}}^{wr}(\tilde{\rho}_{\mathrm{GHZ}}^{wr (sc)})$ are given by
\begin{align}
 &\lambda_1'=\frac{|\beta_1|^2(\bar{p}\bar{r})^n}{\mathcal{T}_2^2}\Big[(2|\alpha_1|^2+|\beta_1|^2p^n)\nonumber\\
 &~~~~~~~~~~~~~~~~~~+\sqrt{4|\alpha_1|^2(|\alpha_1|^2+|\beta_1|^2p^n)}\Big],\nonumber\\
 &\lambda_2'=\frac{|\beta_1|^2(\bar{p}\bar{r})^n}{\mathcal{T}_2^2}\Big[(2|\alpha_1|^2+|\beta_1|^2p^n)\nonumber\\
 &~~~~~~~~~~~~~~~~~~-\sqrt{4|\alpha_1|^2(|\alpha_1|^2+|\beta_1|^2p^n)}\Big],\nonumber\\
 &\lambda_j'=\frac{|\beta_1|^4(p\bar{p}\bar{r})^n}{\mathcal{T}_2^2}, \{j=3,..2^n\},
\end{align}
where $\tilde{\rho}_{\mathrm{GHZ}}^{wr (sc)}$ is the spin conjugated state obtained after applying ${\sigma_y}^{\otimes n}$ on
$\tilde{\rho}_{\mathrm{GHZ}}^{wr*}$.
The $n$--concurrence is, then, given by
\begin{align}
 C_n(\tilde{\rho}_{\mathrm{GHZ}}^{wr}) &= \max{\Big[ 0,\sqrt{\lambda_1'}-\sqrt{\lambda_2'}-\sum_{j=3}^{2^n}\sqrt{\lambda_j'} \Big]}\nonumber\\
 &=  \max \left[ 0, \frac{2|\alpha_1||\beta_1|(\bar{r}\bar{p})^{n/2}}{\mathcal{T}_2}\Big[1-(2^{n-1}-1)\right.\nonumber\\
&~~~~~~~~~~~~~~~~~~~~~~~~~~~~~~~~~~~~~~~~~~~~~~~~\left.\times\frac{|\beta_1|}{|\alpha_1|}p^{n/2}\Big]\right]\nonumber
 \label{conceq}
\end{align}

\begin{align}
 &=\max \Big[ 0,\left(\frac{2|\alpha||\beta|(\bar{s}\bar{r}\bar{p})^{n/2}}{|\alpha|^2\bar{r}^n+|\beta|^2\bar{s}^n(1-pr)^n}\right)\nonumber\\
 &~~~~~~~~~~~~~~~~\cdot\Big[1-(2^{n-1}-1)\frac{|\beta|}{|\alpha|}(\bar{s}p)^{n/2}\Big]\Big].
\end{align}

\begin{figure}[htbp]
\centering
\subfigure[~$n$ = 4]
{
\includegraphics[width=41.25 mm]{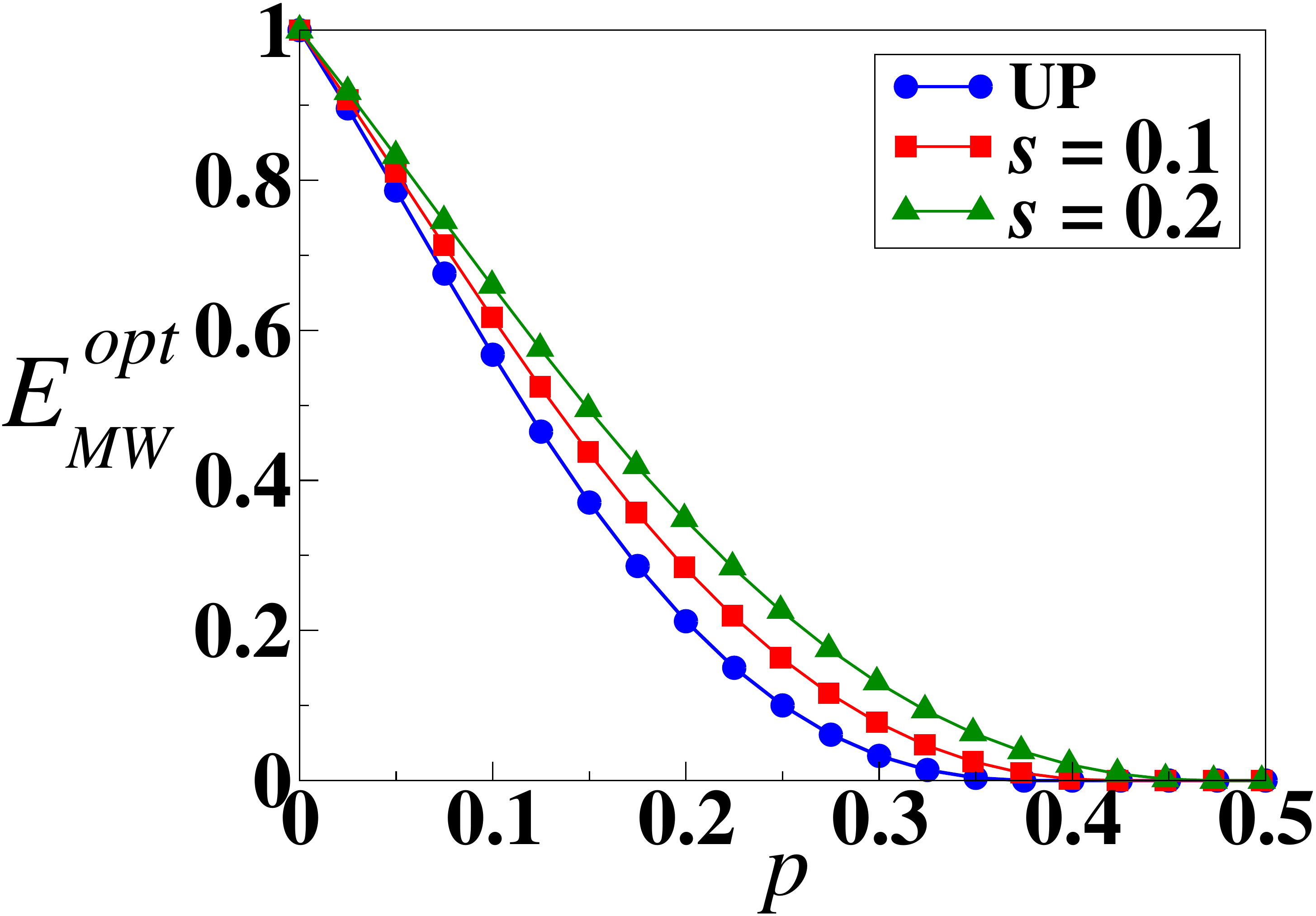}
}
\centering
\subfigure[~$n$ = 8]
{
\includegraphics[width=41.25 mm]{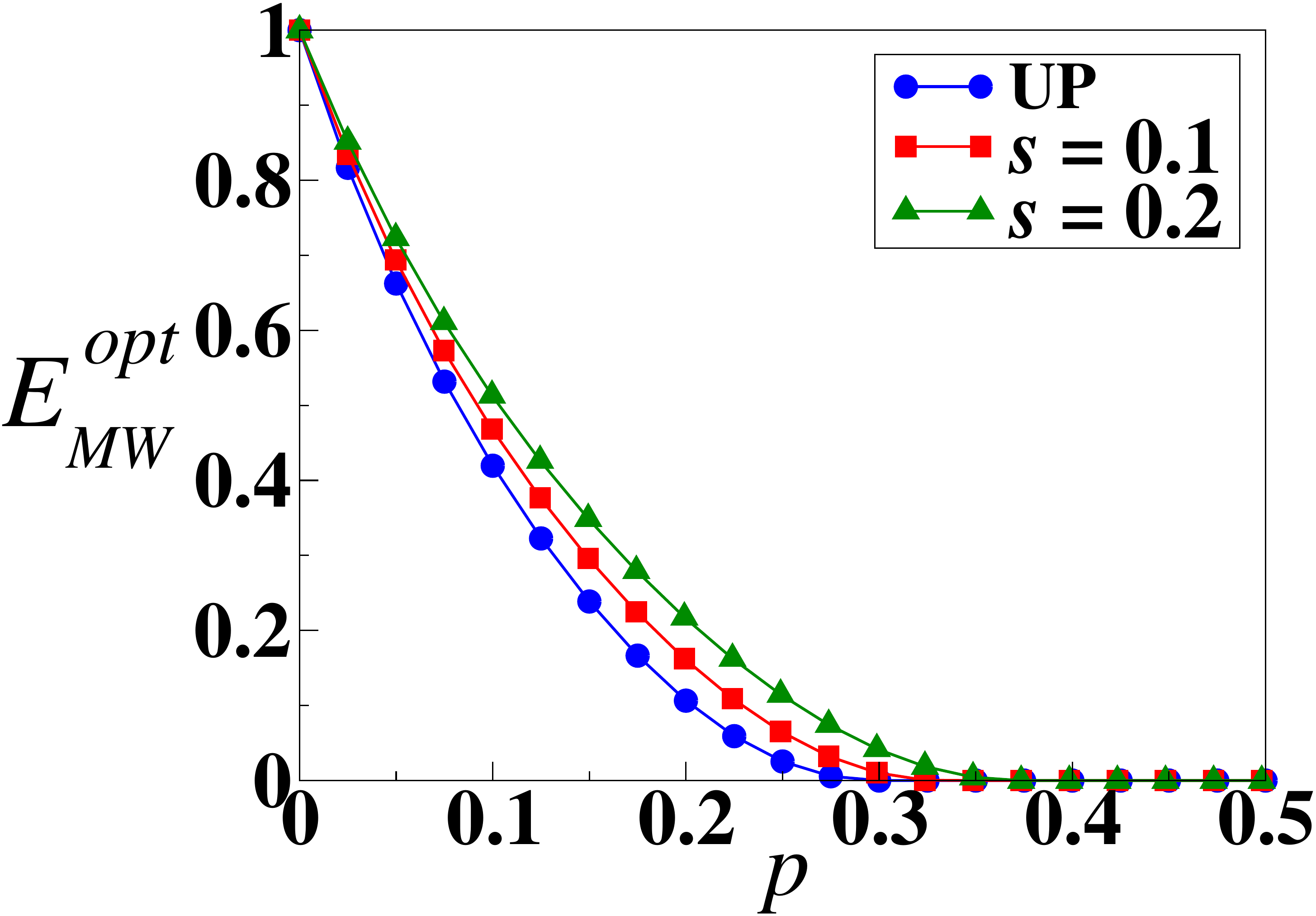}
}
\centering
\subfigure[~$n$ = 12]
{
\includegraphics[width=41.25 mm]{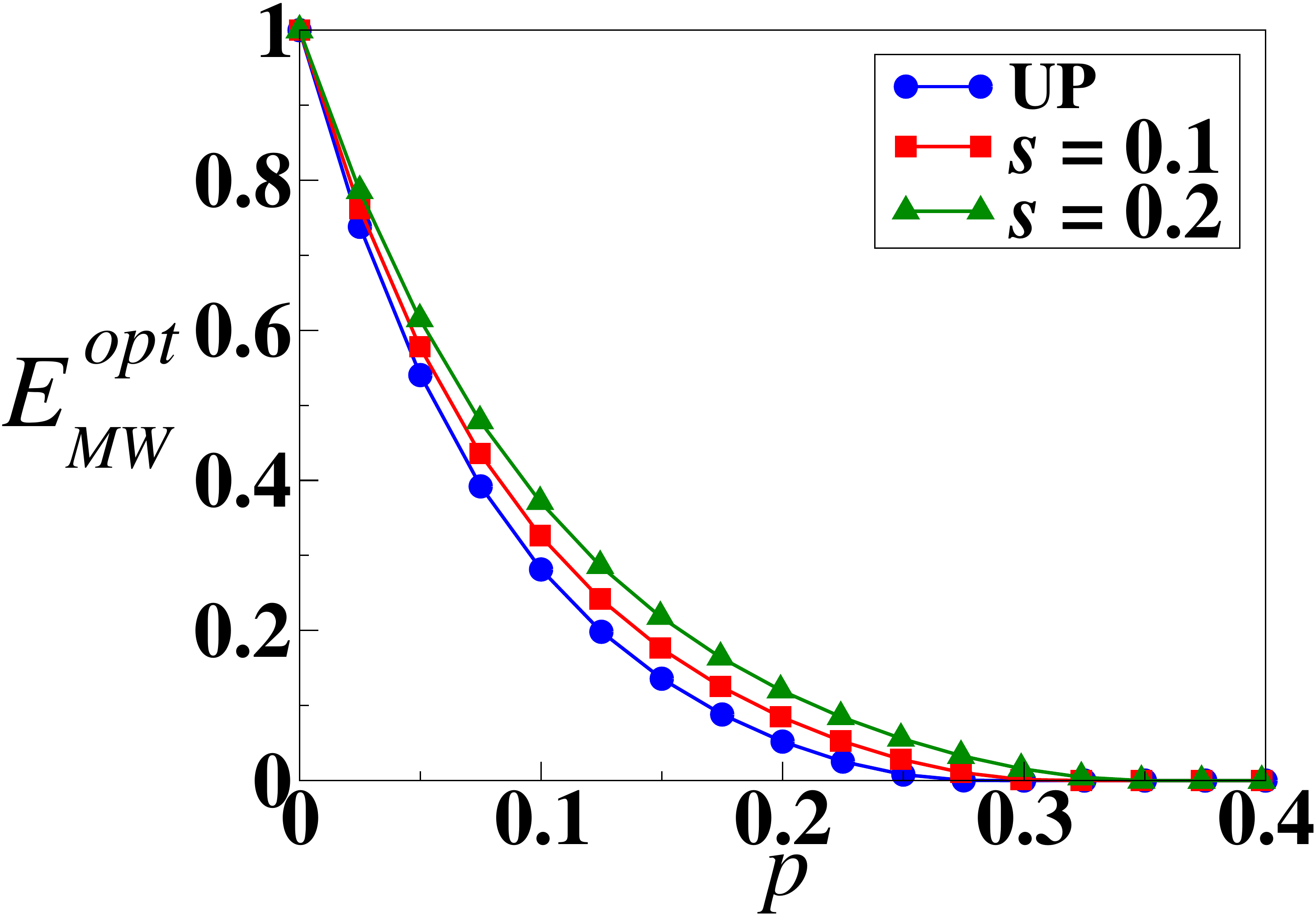}
}
\centering
\subfigure[~$n$ = 24]
{
\includegraphics[width=41.25 mm]{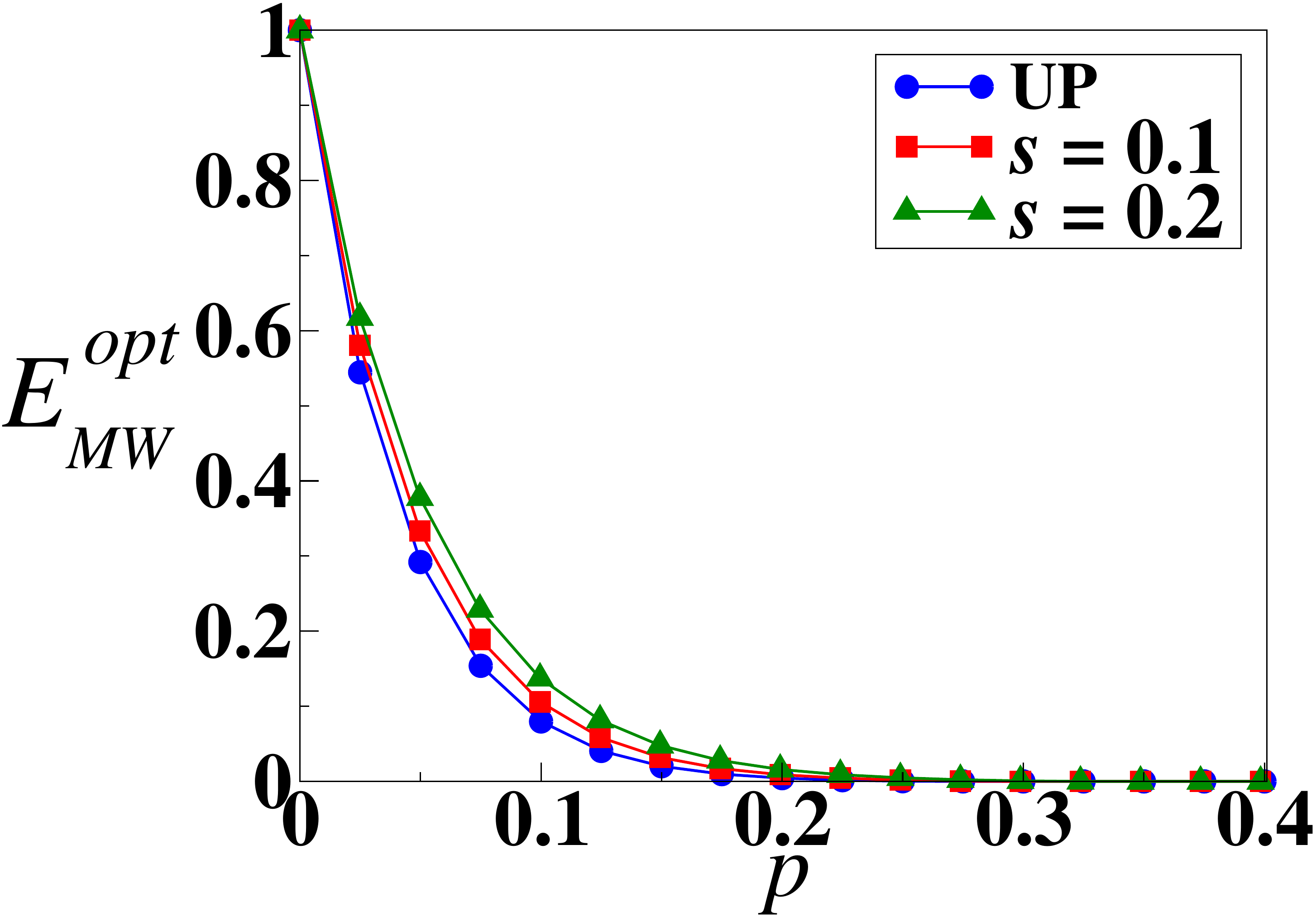}
}
\caption{(Color online) The global entanglement of the gGHZ state with $\alpha =\beta = 1/\sqrt{2}$ as function of
amplitude damping parameter $p$ for different values of initial weak measurement strength $s$ and system size $n$.
We observe that the decay of $E_{MW}^{opt}$ is retarded and the critical value of $p$, $p_c^{MWa}$
is increased for increasing values of $s$. The plot label UP is for the unprotected state ($s$ = $r$ = 0). }
\label{fig3}
\end{figure}

The expression for $C_n(\tilde{\rho}_{\mathrm{GHZ}}^{wr})$ shows that the $n$--concurrence is zero for $p\geq p^{MW}_c$, 
where $p^{MW}_c = \min\{1, \frac{1}{\bar{s}}\left(\frac{|\alpha|}{
|\beta|(2^{n-1}-1)}\right)^{2/n} \}$, and positive definite otherwise. In the limit of very large $n$, the critical value of
$p^{MW}_c$ becomes $1/{(4\bar{s})}$, which is a function of initial weak measurement strength $s$ but independent of the
parameters of the initial gGHZ state. The $n$--concurrence of the protected gGHZ
can sustain decoherence upto $p = 0.25/{\bar{s}}$ for very large $n$. This indicates that the protected
state is more robust against decoherence and ESD occurs at higher amplitude damping values, as compared to the unprotected state.  

The global entanglement, $E_{MW}$ is given by
$C_n^2(\tilde{\rho}_{\mathrm{GHZ}}^{wr})$.
The optimal value of the global entanglement, $E_{MW}^{opt}$, after the weak measurement reversal
protocol can be obtained by numerically maximizing $E_{MW}$, with respect to the weak reversal strength \(r\), at
fixed values of \(n\), \(s\), and \(p\). Fig. \ref{fig3} shows the optimal value of global entanglement, $E_{MW}^{opt}$,
against the amplitude damping
parameter \(p\), for different values of
\(n\) and \(s\). The plots show that $E_{MW}^{opt}$ increases with weak
measurement strength \(s\).
Again, arbitrary large values of $s$ in $[0,1]$ are limited, as it will lead to low success probability of the weak
measurement reversal protocol.
It is also clear from the plots that the critical value \(p^{MWa}_c\) is always greater for
the weak measurement applied state compared to the unprotected state.
We observe that the value of \(p^{MWa}_c\) is not suitably enhanced for very large $n$.
%
\begin{figure}[htbp]
\centering
\includegraphics[width=50 mm]{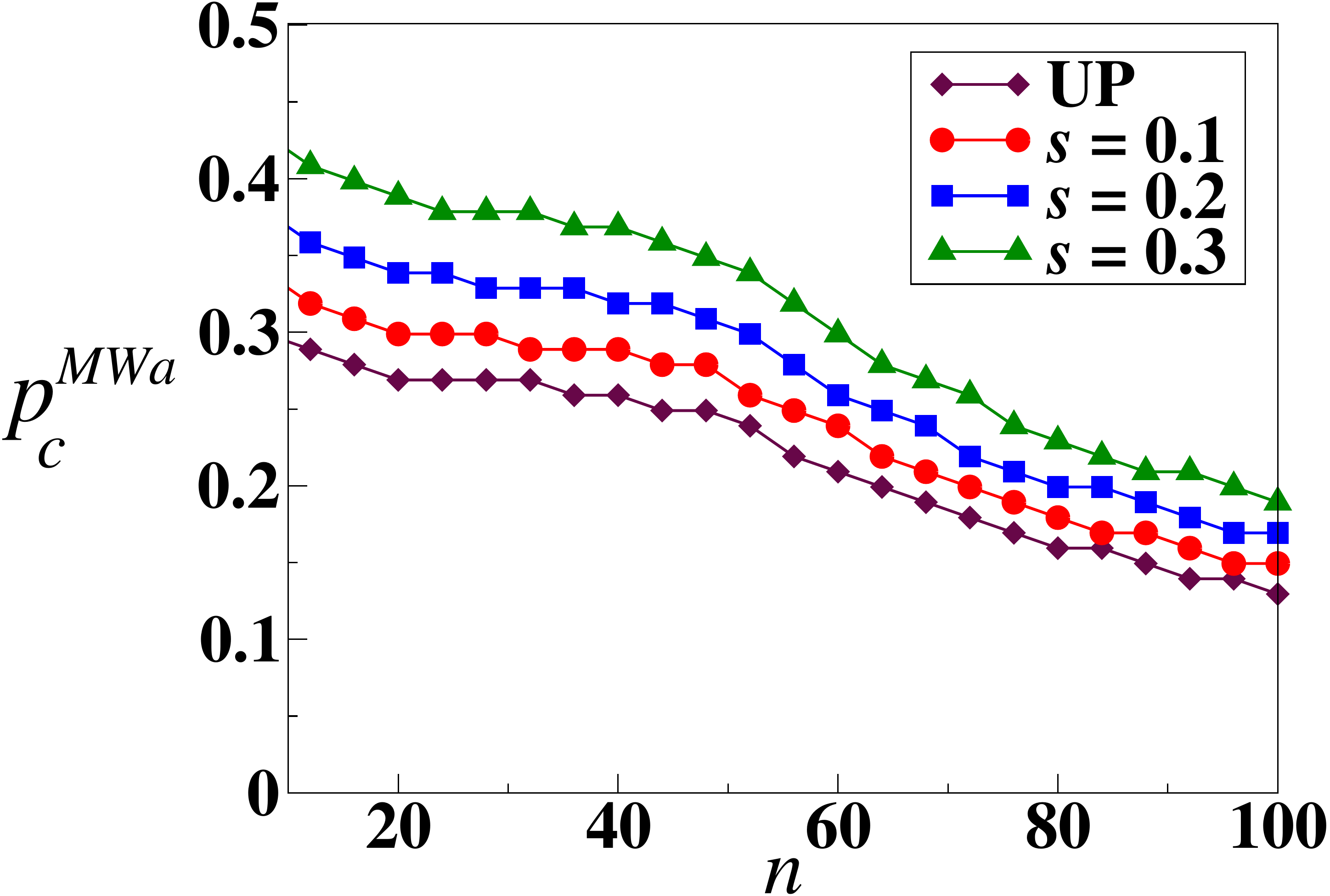}
\caption{(Color online) The critical damping value, $p=p^{MWa}_c$, for $E_{MW}^{opt}$ is plotted against the number of qubits $n$. 
The  plot shows that the critical value, $p^{MWa}_c$, of the protected state remains above the unprotected state for $n$ = 100, though
the difference diminishes with increasing $n$. The plot label UP is for the unprotected state ($s$ = $r$ = 0).}
\label{fig:fa}
\end{figure}

\section{Quantum teleportation in one sender, many receivers setting and multiparty quantum information splitting}
\label{telep}
In Section (\ref{res}), we have seen that multipartite entanglement can be protected against LADC, using weak measurement
reversal protocol. The immediate question that arises, is whether such a protection scheme can prove beneficial
in some quantum communication task? Here we consider, two proximately related quantum communication tasks, namely, quantum
teleportation of an unknown qubit state in a one sender, many receivers setting \cite{teleorig, hill} and
multiparty quantum information splitting \cite{hill,split}, that, under decoherence, are known to be associated with different
aspects of multiparty quantum correlations \cite{rafael}.
We consider the decohered gGHZ state as the shared resource in these quantum communication tasks and calculate the average fidelities under the weak measurement protection protocol.
%

According to the one sender, many receivers quantum teleportation protocol,
Alice ($A$) wants to send an unknown qubit in a state ($\ket{\psi_0}=\alpha\ket{0}+\beta\ket{1}$) to
($n-1$) Bobs ($B_i$, $i = 1$ to $n-1$) using a shared $n$--qubit GHZ state, given as
\begin{eqnarray}
 \ket{\phi}_{A:B_1..B_{n-1}}=\frac{1}{\sqrt{2}}[\ket{0}^{\otimes n}+\ket{1}^{\otimes n}].
\label{sghz}
\end{eqnarray}
The success of a quantum teleportation protocol can be evaluated by the average fidelity between the unknown initial qubit
to be teleported
and the final qubit received during the process. The fidelity of teleportation is defined as $\cal{F}=\bra{\psi_0} \rho \ket{\psi_0}$, where
$\ket{\psi_0}$ is the unknown initial state and $\rho$ is the final teleported state. For a perfect quantum teleportation protocol,
with unit fidelity, $A$ and $B_i$'s must share a maximally quantum correlated state \cite{teleorig,hortel}. 
The maximum fidelity achievable using only
classical communication, with no quantum correlations, is $\cal{F}_{cl}$ = $2/3$ \cite{clone}. 

For the considered $n$--qubit GHZ state, given by (\ref{sghz}), the pre-shared amount of quantum correlation is sufficient to teleport the unknown state 
from Alice to ($n-1$) Bobs with maximum fidelity. 
To elaborate, Alice makes a Bell measurement on the unknown qubit, $\ket{\psi_0}$, and the qubit ($A$) of the $n$--qubit
pre-shared GHZ state, in her possession, and classically communicates her results to ($n-1$) Bobs. The information to be
teleported is encoded in the reduced ($n-1$)-qubit state in Bobs possession.
On receiving Alice's result, the ($n-1$) Bobs together apply suitable local unitaries to obtain the state 
\begin{equation}
\ket{\psi_f}=\alpha|0\rangle^{\otimes (n-1)}+\beta|1\rangle^{\otimes (n-1)}.
\label{dis}
\end{equation} 
Applying local measurements, the ($n-1$) Bobs together can distill the unknown state, $\ket{\psi_0}=\alpha\ket{0}+\beta\ket{1}$,
with unit fidelity, from the encoded state, $\ket{\psi_f}$.

However, for practical applications, the pre-shared resource is generally not optimal for perfect teleportation.
Let us consider a slightly non-idealistic situation where some arbitrary quantum state preparator prepares the $n$--qubit gGHZ
state, given by Eq. (\ref{sghz}), and sends the qubits to Alice and Bobs.
Since Alice and Bobs are at distant locations, the state sent by the preparator suffers from decoherence due to environmental interaction.
Such a decohered state can be modelled by sending the prepared $n$--qubit gGHZ state through $n$ single--qubit LADCs (Sec.\ref{amp}) to Alice and Bobs.
%
%
Thus, the state shared by Alice and $(n-1)$ Bobs is given in Eq. ({\ref{gdamp})} (with $\alpha=\beta=1/\sqrt{2}$).
The average fidelity of teleportation using such a shared state
is known to decrease with increasing damping strength $p$ \cite{rafael}. 

In the weak measurement reversal protocol, the preparator first applies a weak measurement of strength \(s\) on each shared qubit
of the state. These weak measured states are then 
sent to Alice and ($n-1$) Bobs through the  local amplitude damping channel, with strength $p$. Now, Alice and ($n-1$) Bobs locally
apply reversal weak  measurements of equal strength \(r\), to all the $n$ qubits. 
The final measured state, using Eq. (\ref{proghz}), is now the pre-shared quantum state, given by
\begin{align}
\rho_{A:B_1..B_{n-1}}^{wr} &=\frac{1}{2\mathcal{T}}\big[|\bar{r}^n(\ket{0}\bra{0})^{\otimes n}+(\bar{r}\bar{p}\bar{s})^{n/2}
[(\ket{0}\bra{1})^{\otimes n}\nonumber\\
&+ h.c.] +\bar{s}^{n}\sum_{k=0}^{n}(p\bar{r})^k\bar{p}^{(n-k)}[(\ket{0}\bra{0})^{\otimes k} \nonumber\\
&\otimes(\ket{1}\bra{1})^{\otimes n-k}+ \cal{R}_1]\Big],
\label{tel-res}
\end{align}
Alice now wants to teleport the unknown qubit $\ket{\psi_0}$ to ($n-1$) Bobs using the $n$--qubit pre-shared
state $\rho_{A:B_1..B_{n-1}}^{wr}$. The joint state is given by,
$
\ket{\psi_0}\bra{\psi_0}\otimes\rho_{A:B_1..B_{n-1}}^{wr}.
$
Alice now possesses a qubit ($A$) of the $n$--qubit pre-shared state $\rho_{A:B_1..B_{n-1}}^{wr}$ and the unknown qubit
$\ket{\psi_0}$ and
makes a Bell basis measurement on the two qubits in her possession. 
The unknown state to be teleported is now encoded in some reduced ($n-1$)--qubit state in ($n-1$) Bobs possession.
Let, $\ket{\phi^\pm}=\frac{1}{\sqrt{2}}(\ket{00}\pm \ket{11})$ and $\ket{\psi^\pm}=\frac{1}{\sqrt{2}}(\ket{01}\pm \ket{10})$, be the Bell basis states.
On measurement, 
Alice gets $\ket{\phi^\pm}$ as outcomes with probability $p_1^{\pm}$ and 
the normalized state of ($n-1$) Bobs is $\rho_1^{\pm}$, where
\begin{align}
\rho_1^{\pm}&=\frac{1}{4p_1^{\pm}\cal{T}}\Big[|\alpha|^2\bar{r}^n\ket{0}\bra{0}^{\otimes(n-1)}\nonumber\\
&+\bar{s}^n(|\alpha|^2p\bar{r}+|\beta|^2\bar{p})\big(p\bar{r}\ket{0}\bra{0}+
\bar{p}\ket{1}\bra{1}\big)^{\otimes(n-1)}\nonumber\\
&\pm(\bar{r}\bar{p}\bar{s})^{n/2}\big\{\alpha\beta^*\ket{0}\bra{1}^{\otimes(n-1)}+h.c.\big\}\Big]_{B_1..B_{n-1}},
\label{rho1}
\end{align}
and
\begin{align}
 p_1^+ = p_1^-= \frac{|\alpha|^2\bar{r}^n + \bar{s}^n(|\alpha|^2p\bar{r}+|\beta|^2\bar{p})(p\bar{r}+\bar{p})^{n-1}}{4\mathcal{T}} .
\label{p1}
\end{align}
Similarly, Alice gets $\ket{\psi^\pm}$ as outcome with probability $p_2^{\pm}$ and the
normalized state of ($n-1$) Bobs is $\rho_2^{\pm}$, where
\begin{align}
\rho_2^{\pm}&=\frac{1}{4p_2^{\pm}\cal{T}}\Big[|\beta|^2\bar{r}^n\ket{0}\bra{0}^{\otimes(n-1)}\nonumber\\
&+\bar{s}^n(|\beta|^2p\bar{r}+|\alpha|^2\bar{p})\big(p\bar{r}\ket{0}\bra{0}+
\bar{p}\ket{1}\bra{1}\big)^{\otimes(n-1)}\nonumber\\
&\pm(\bar{r}\bar{p}\bar{s})^{n/2}\big\{\alpha\beta^*\ket{1}\bra{0}^{\otimes(n-1)}+h.c.\big\}\Big]_{B_1..B_{n-1}},
\label{rho2}
\end{align}
and
\begin{align}
 p_2^+ = p_2^-= \frac{|\beta|^2\bar{r}^n + \bar{s}^n(|\beta|^2p\bar{r}+|\alpha|^2\bar{p})(p\bar{r}+\bar{p})^{n-1}}{4\mathcal{T}} .
\label{p2}
\end{align}
Now depending on the measurement outcomes of Alice, the ($n-1$) Bobs apply suitable local unitaries on the encoded state,
${\rho_i}^{\pm}$ ($i=1, 2$), to obtain the optimal ($n-1$)--qubit state, say, ${\tilde{\rho}_j}$ ($j=1~\mathrm{to}~4$), with
corresponding probabilities
$\tilde{p}_j$ = \{$p_1^{\pm}$, $p_2^{\pm}$\}, that maximizes the fidelity of the teleportation. 
%
%
%
From relation (\ref{dis}), it is known that in the absence of LADC, the encoded state $|\psi_f\rangle$, allows ($n-1$) Bobs
to distill the unknown state, $|\psi_0\rangle$, with unit fidelity. 
Hence, the success of the protocol, depends on the fidelity of ($n-1$) Bobs optimal encoded state, $\tilde{\rho_j}$, with
the state $|\psi_f\rangle$.
Thus, the average fidelity of teleportation can be calculated using the relation,
$\langle\cal{F}\rangle_{tel} = \sum_j \tilde{p}_j\langle\psi_f|\tilde{\rho_j}|\psi_f\rangle$.
%
%
%

\begin{figure}[htbp]
\centering
\subfigure[~$n$ = 4]
{
\includegraphics[width=41.25 mm]{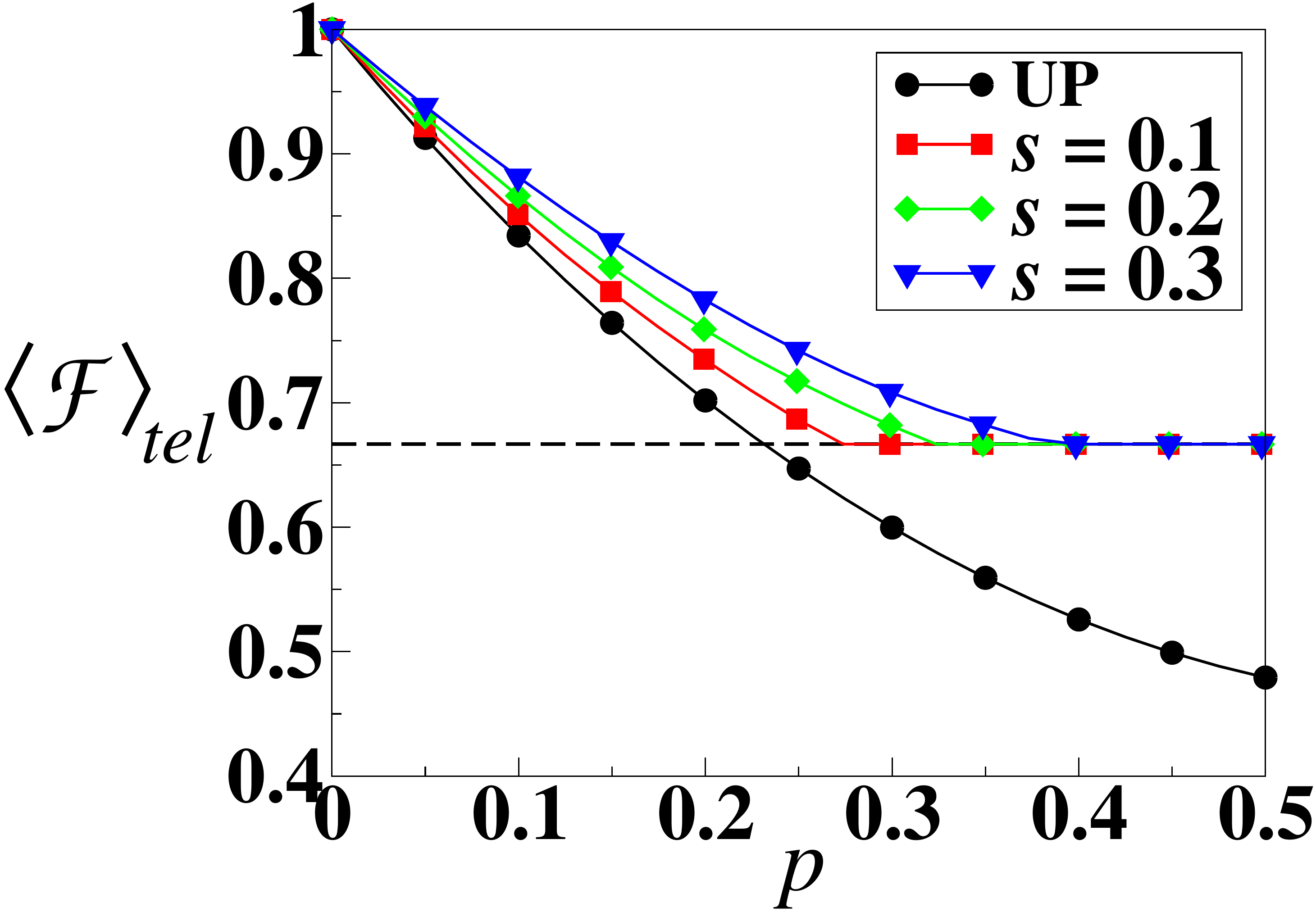}
}
\centering
\subfigure[~$n$ = 8]
{
\includegraphics[width=41.25 mm]{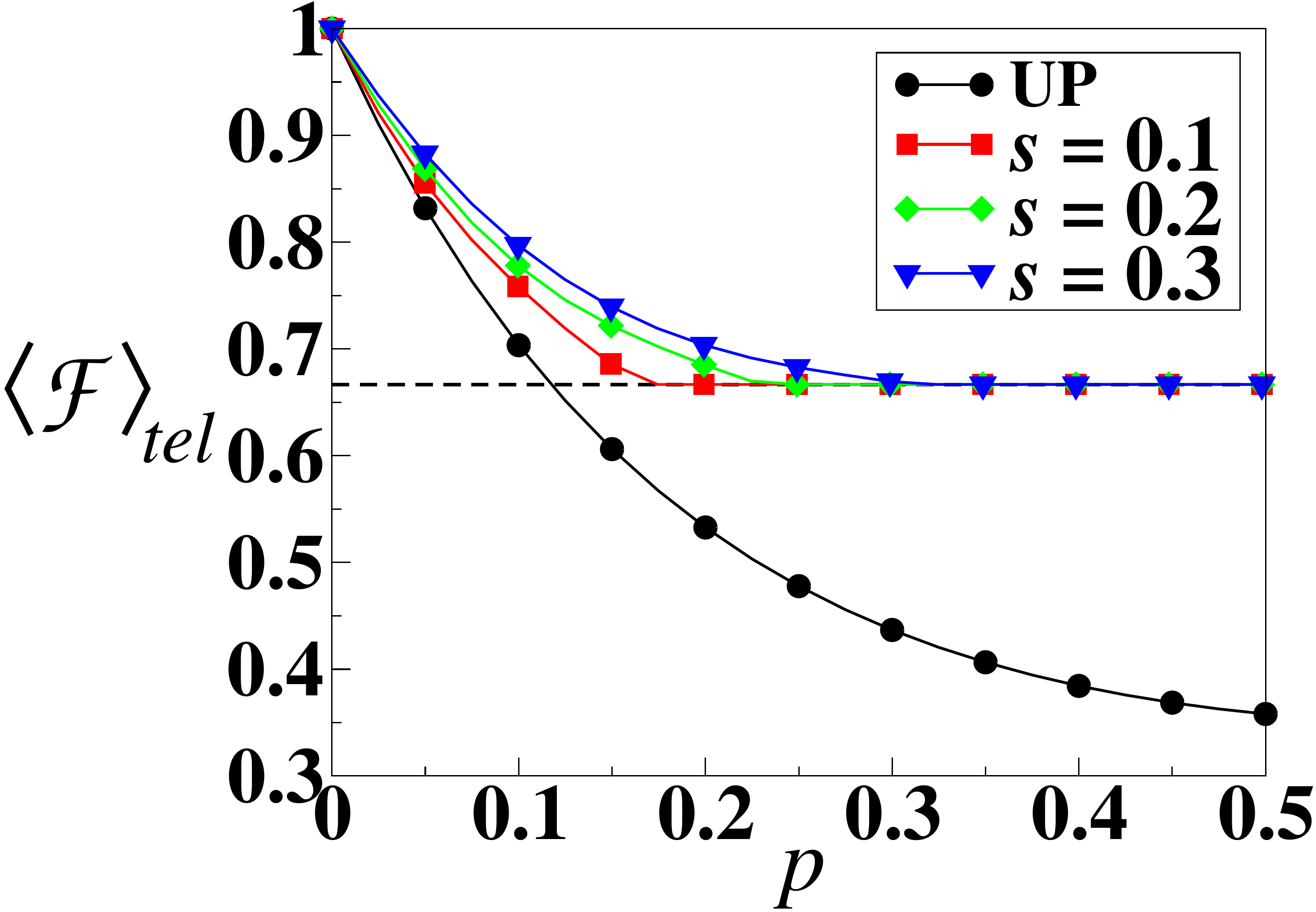}
}
\centering
\subfigure[~$n$ = 12]
{
\includegraphics[width=41.25 mm]{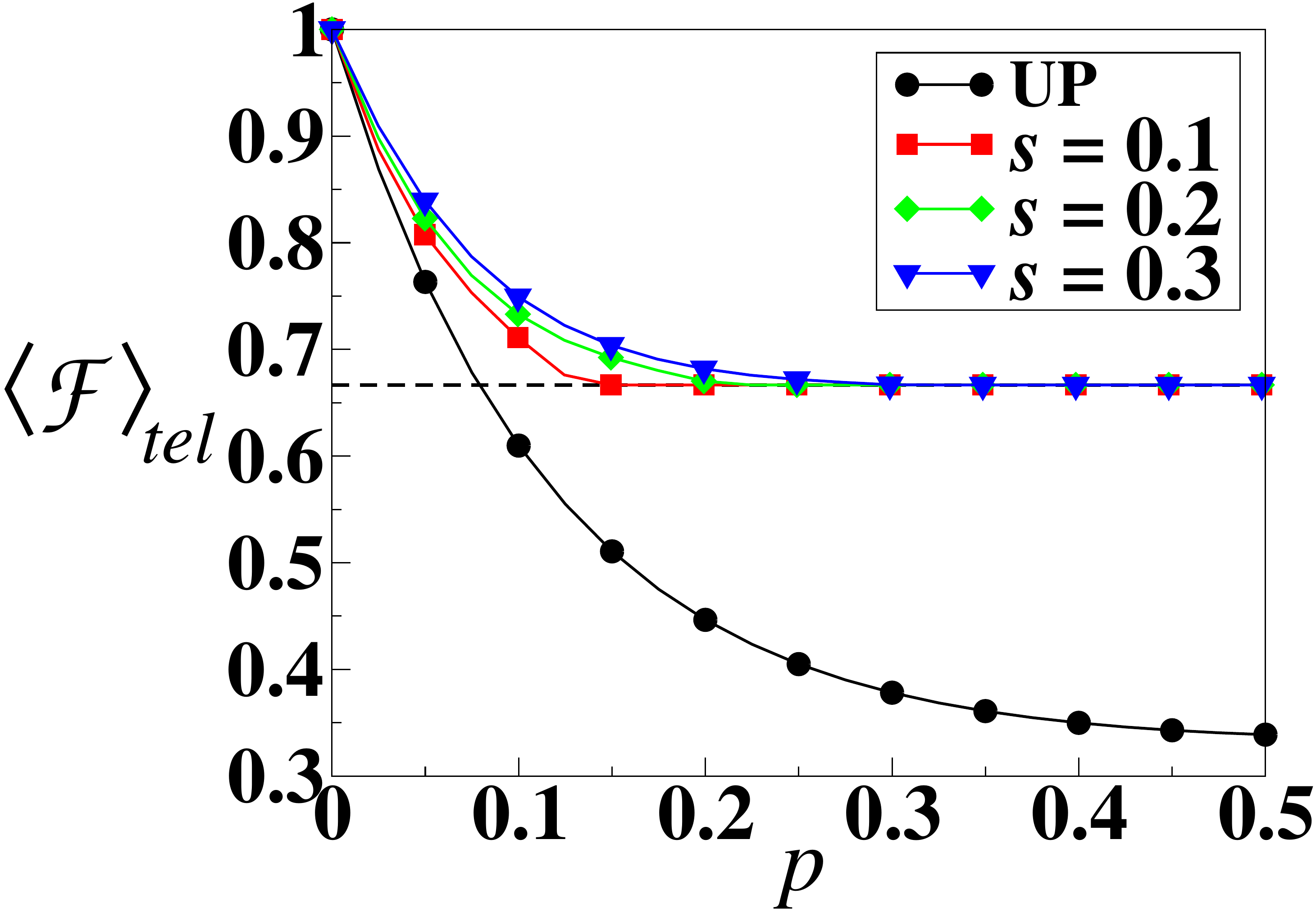}
}
\centering
\subfigure[~$n$ = 24]
{
\includegraphics[width=41.25 mm]{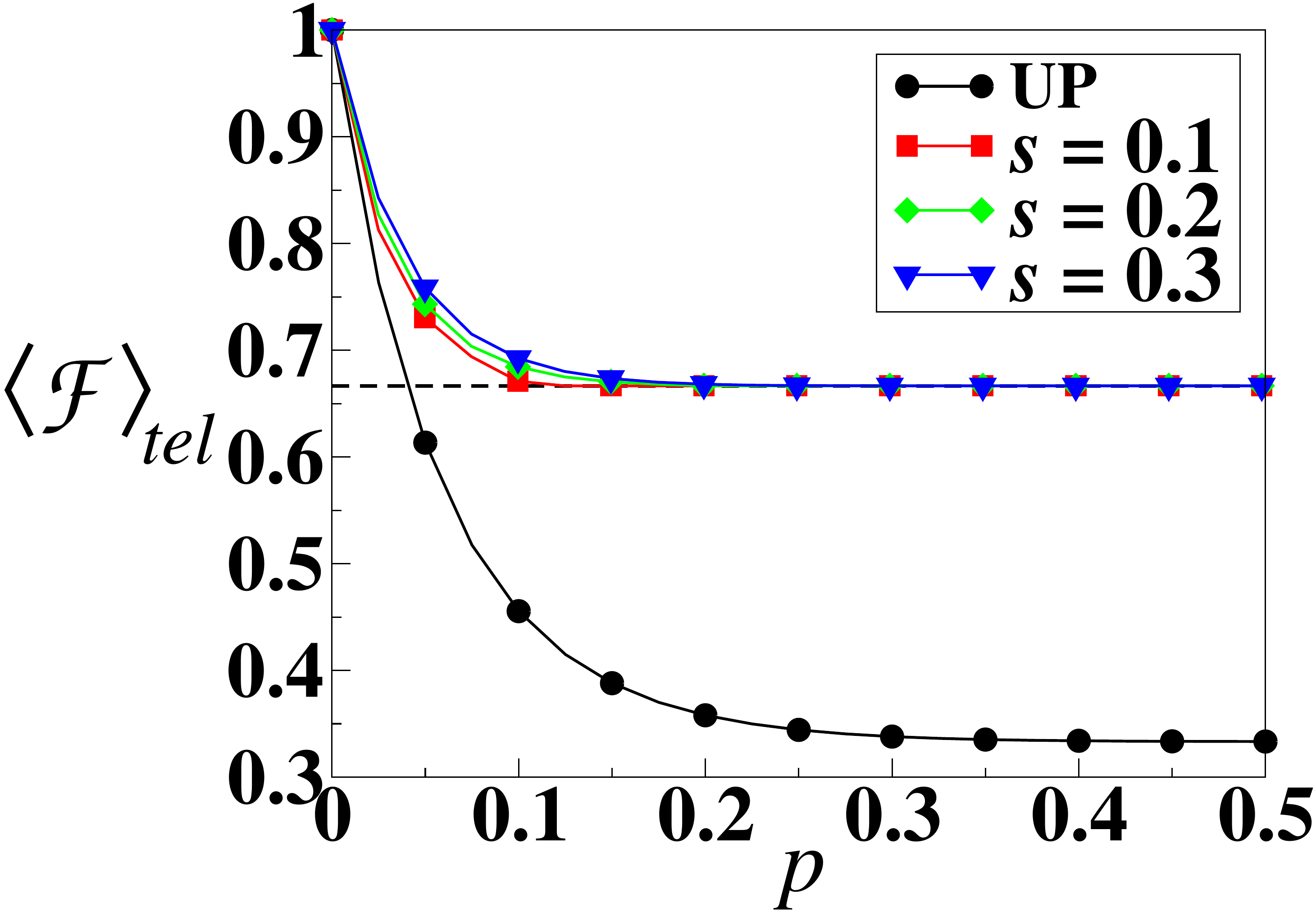}
}
\caption{(Color online) The average fidelity of teleportation ($\langle\cal{F}\rangle_{tel}$) as a function of amplitude damping parameter
$p$ for different values of initial weak measurement strength $s$ and number of qubits $n$. We observe that the noisy channel under
weak measurement reversal protocol never reduces below the classical resource limit by always generating an average fidelity,
$\langle\cal{F}\rangle_{tel}\ge 2/3$. We also observe that the critical damping value leading to the classical limit,
under amplitude-damping, is increased for finite $s$. At $p=1$, the average fidelity of teleportation for unprotected state
becomes $2/3$, which we have not shown for the sake of brevity. The plot label UP is for the unprotected state ($s$ = $r$ = 0). }
\label{fig5}
\end{figure}

Using the weak measurement reversal protocol, the average fidelity ($\langle\cal{F}\rangle_{tel}$) of quantum teleportation is
given by the expression
\begin{align}
\langle\cal{F}\rangle_{tel}=&\frac{1}{6\mathcal{T}}\big[2\bar{r}^n(1+p^n\bar{s}^n)+2\bar{p}^n\bar{s}^n\nonumber\\
 &+\bar{s}^n(p\bar{r}\bar{p}^{n-1}+\bar{p}(p\bar{r})^{n-1})+2(\bar{r}\bar{p}\bar{s})^{n/2}\big].
\end{align}
The average fidelity of teleportation is maximized with respect to the  reversal weak measurement strength $r$, keeping the
parameters $p$, $s$, and $n$ fixed.
Without the protection protocol ($r = 0$ and $s=0$), the average fidelity is given by the expression 
$(2+p^{n-1}(1+p)+\bar{p}^{n-1}(1+\bar{p})+2\bar{p}^{n/2})/6$, as obtained in \cite{rafael}.
%
Figure (\ref{fig5}) is the plot for the numerically optimized average fidelity of quantum teleportation. It can be observed from the
figure that the critical value $p_c^f(n)$ (beyond which the optimized average fidelity is less than or equal to 
the classical teleportation fidelity, $2/3$)
has increased with the initial weak measurement strength $s$, as compared to the average fidelity without any weak measurement protection.
It is also observed from Fig. (\ref{fig5}) that the optimal average fidelity always remains above or equal to classical fidelity
that is $\langle\cal{F}\rangle_{tel} \ge \cal{F}_{cl}$, for finite weak measurement reversal application.

In the one sender, many receivers quantum teleportation protocol, the average fidelity of teleportation is obtained by
maximizing the fidelity of the encoded ($n-1$)--qubit states in possession of Bobs with the state $|\psi_f\rangle$.
The ($n-1$) Bobs perform as a single entity to optimize the outcome. 
An alternate approach is to consider the multiparty quantum information splitting protocol \cite{hill,split} to
decode the unknown state, $|\psi_0\rangle$. In this protocol, the ($n-1$) Bobs are independent, but co-operative,
entities that can classically communicate with each other. A specific Bob (say ($n-1$)$^{th}$ Bob) is chosen as
the pre-determined receiver with the other ($n-2$) Bobs assisting it to receive the unknown state, $|\psi_0\rangle$.
The information is thus being split among many receivers and can be retrieved only by necessary co-operation. This forms
the basis of many secret sharing protocols \cite{hill}. 

In the quantum information splitting protocol, the encoding of the state to be teleported by Alice remains the same as
the quantum teleportation protocol. Alice makes a Bell basis measurement on the two qubits in her possession, the qubit ($A$)
of the $n$--qubit pre-shared state $\rho_{A:B_1..B_{n-1}}^{wr}$ and the unknown qubit $\ket{\psi_0}$. 
%
After Alice's measurement, the encoded states, ${\rho_j} = \{{\rho_1}^{\pm}, {\rho_2}^{\pm}\}$ ($j=1$ to $4$), given by Eq.(\ref{rho1}) and (\ref{rho2}),
with corresponding probabilities $p_j = \{p_1^{\pm}, p_2^{\pm}\}$, are split among the independent ($n-1$) Bobs. The ($n-2$) non-receiver Bobs measure their
respective qubits in the eigenbasis of the Pauli spin matrix $\sigma_x$ and classically communicate their outcomes, which
occur with probabilities,
$q^j_k$ ($j=1~\mathrm{to}~4,~k=1~\mathrm{to}~2^{n-2}$), to the ($n-1$)$^{th}$ Bob. Depending on the measurement outcomes of
Alice and the ($n-2$) Bobs,
($n-1$)$^{th}$ Bob applies suitable unitary operations to obtain the single--qubit state, 
say, $\tilde{\sigma}^j_k$ ($j=1~\mathrm{to}~4,~k=1~\mathrm{to}~2^{n-2}$),
with probability $p_jq^j_k$. The average fidelity of the protocol can then be calculated using the relation,
$\langle\cal{F}\rangle_{is} =  \sum_{j,k}  p_jq^j_k\langle\psi_0|\tilde{\sigma}^j_k|\psi_0\rangle$. For a maximally entangled pre-shared
state, such as the $n$--qubit GHZ state, the ($n-1$)$^{th}$ Bob can obtain the unknown state, $\ket{\psi_0}$, with unit fidelity.

\begin{figure}[htbp]
\centering
\subfigure[~$n$ = 4]
{
\includegraphics[width=41.25 mm]{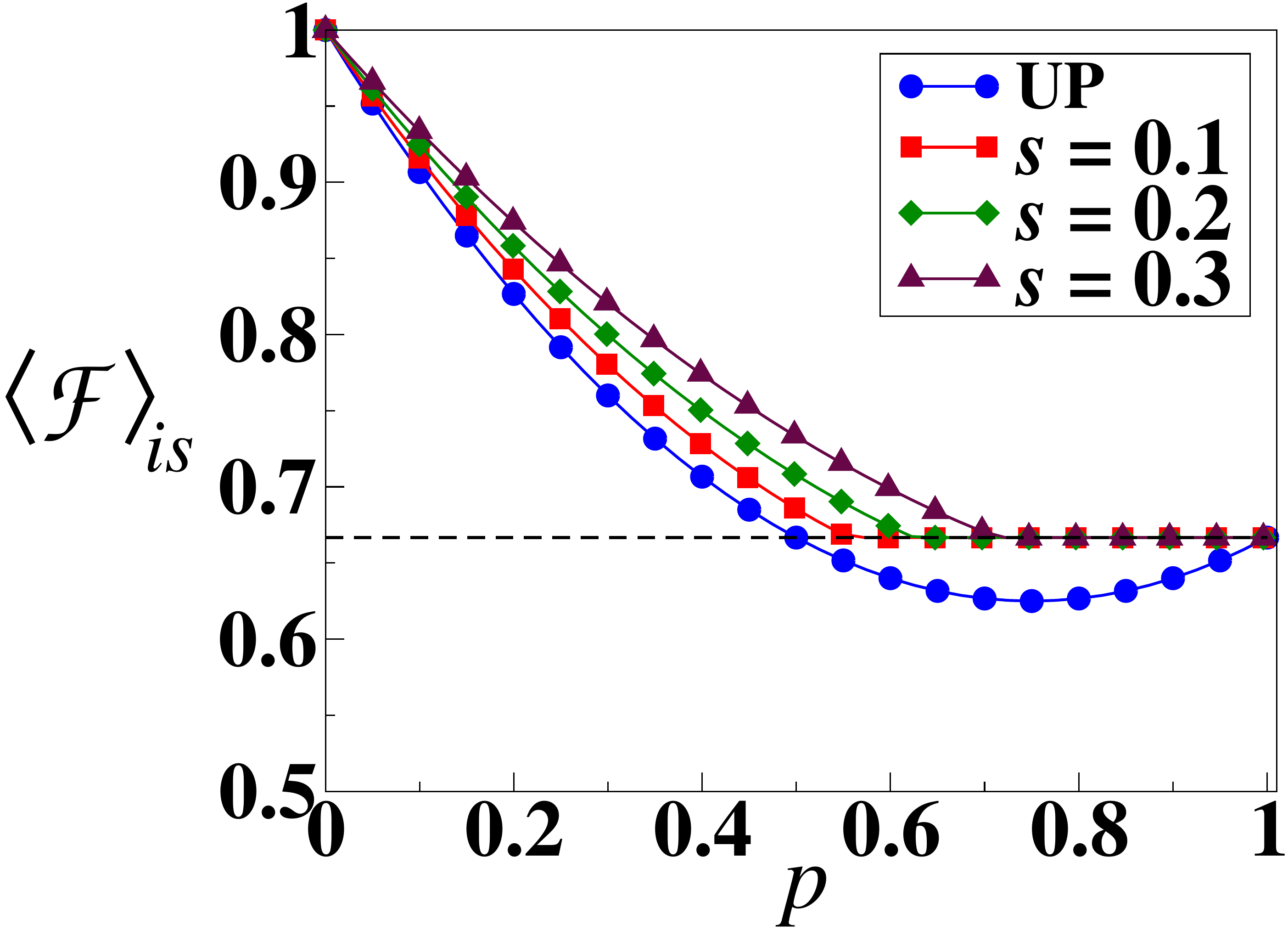}
}
\centering
\subfigure[~$n$ = 8]
{
\includegraphics[width=41.25 mm]{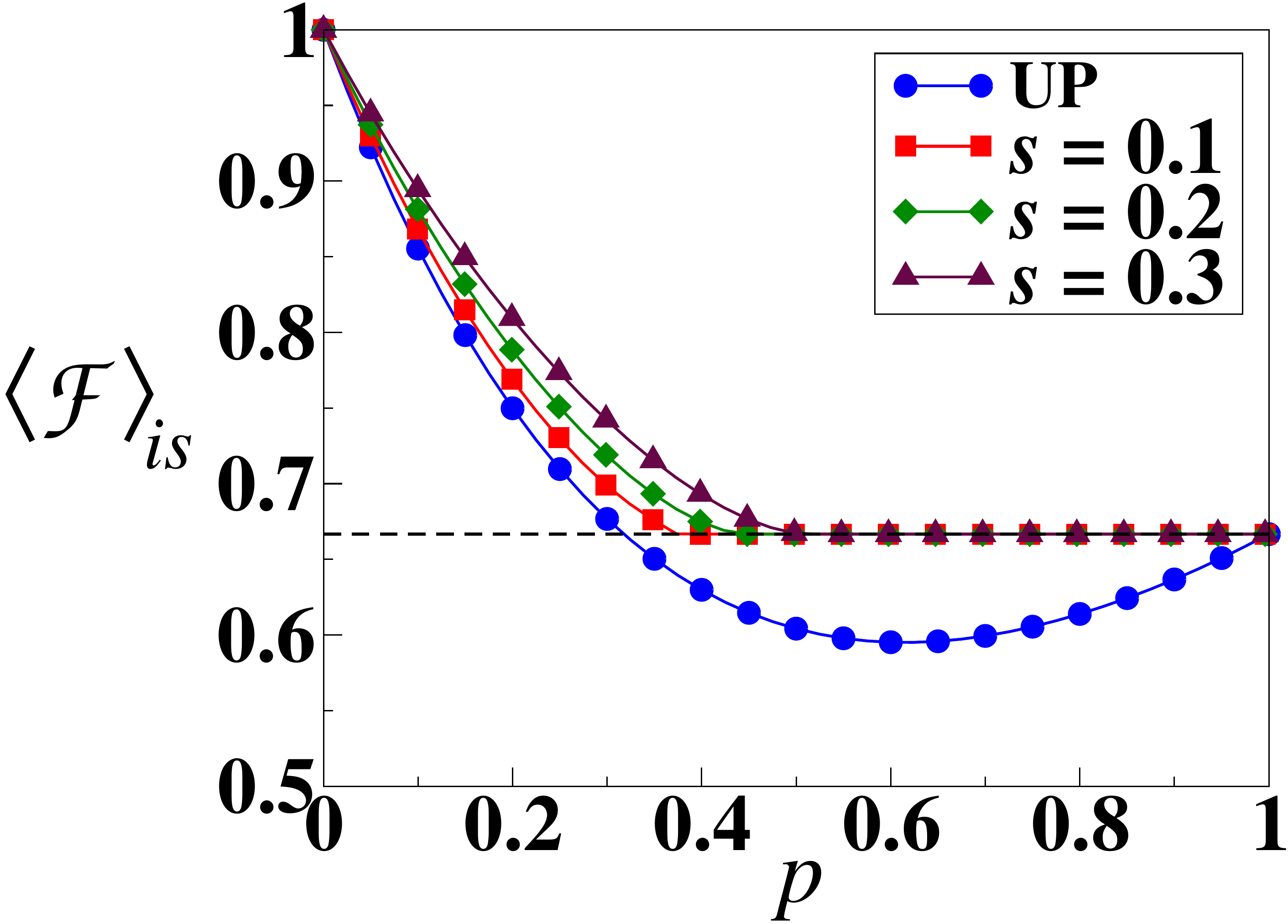}
}
\centering
\subfigure[~$n$ = 12]
{
\includegraphics[width=41.25 mm]{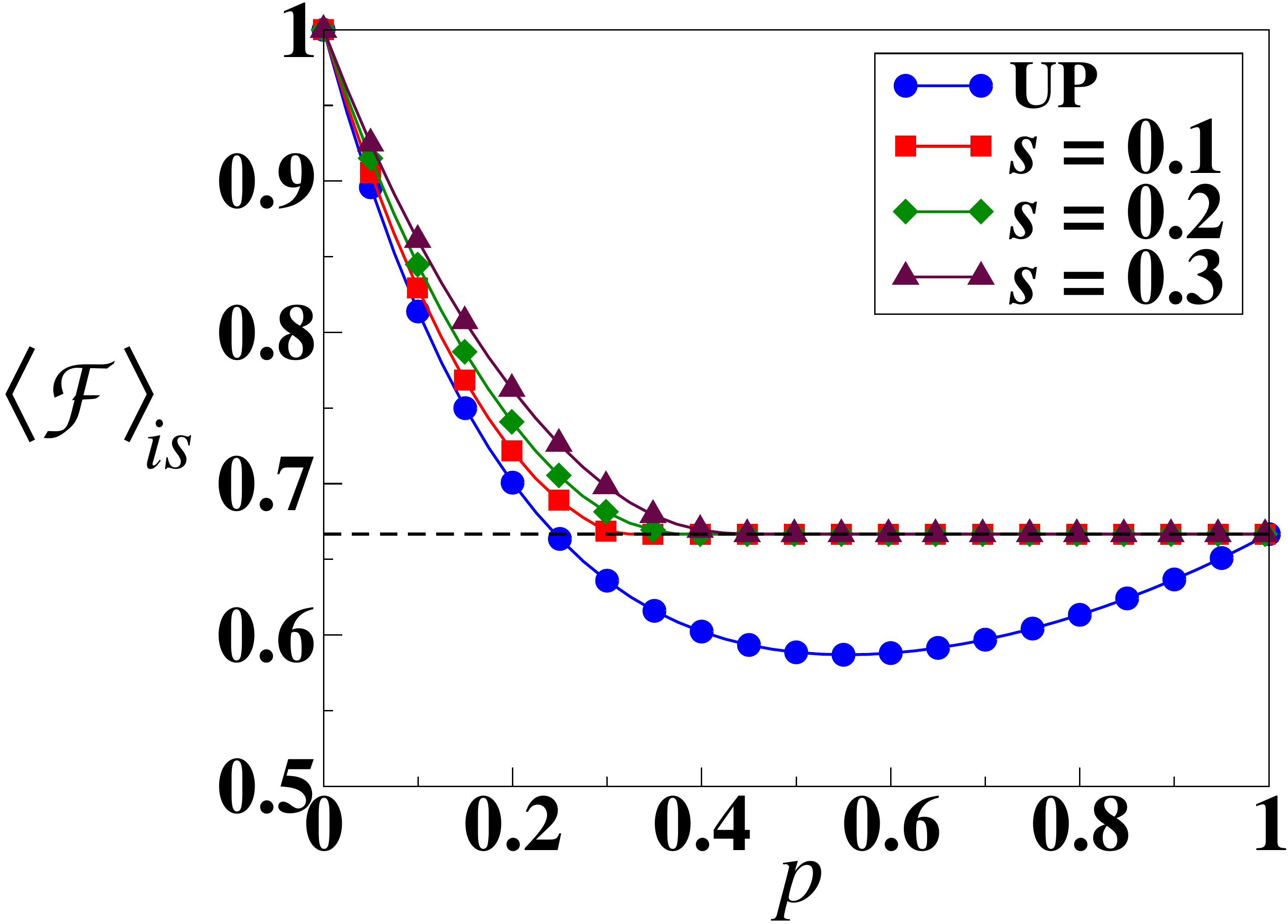}
}
\centering
\subfigure[~$n$ = 24]
{
\includegraphics[width=41.25 mm]{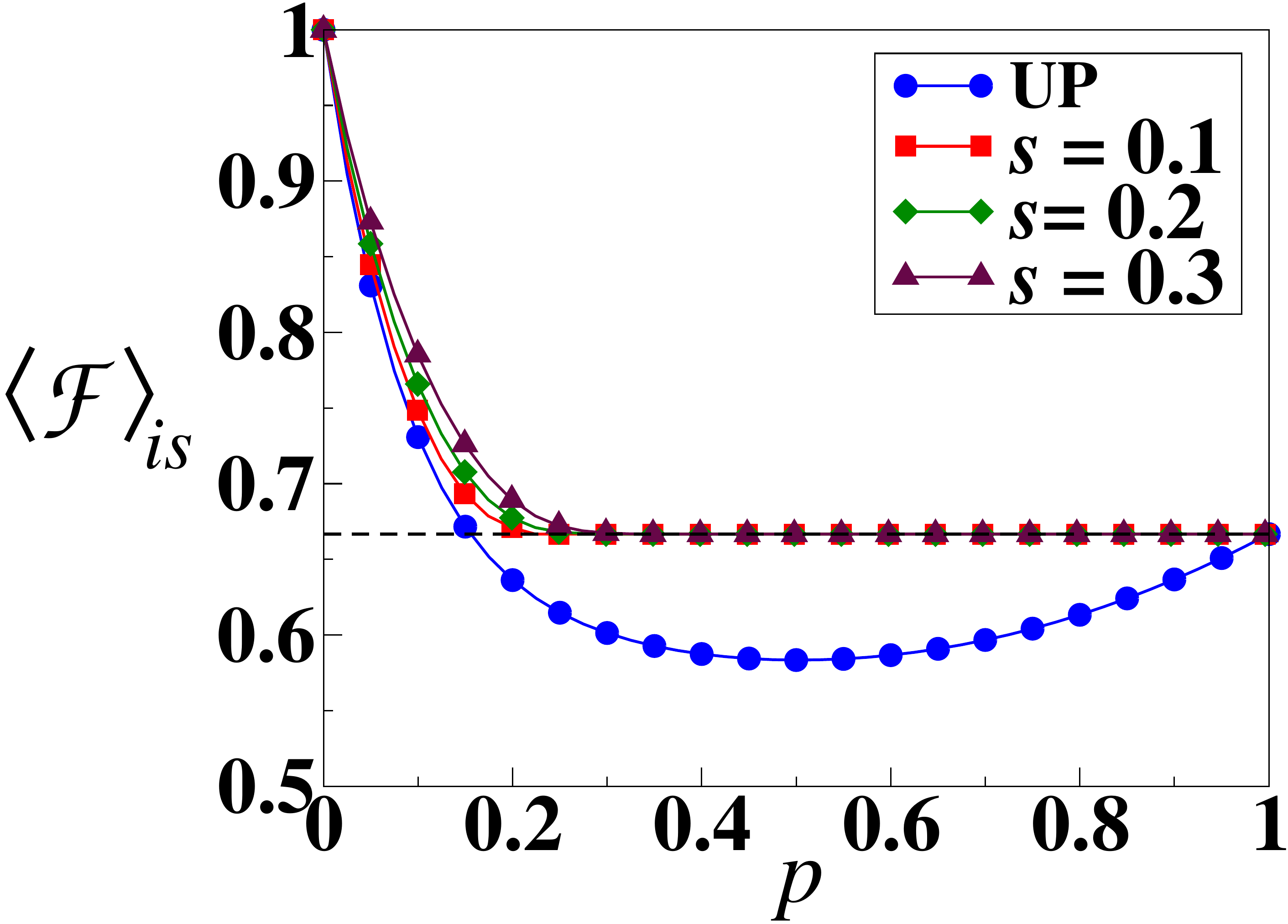}
}
\caption{(Color online)The average fidelity of teleportation using information splitting ($\langle\cal{F}\rangle_{is}$) as a function of amplitude
damping parameter
$p$ for different values of initial weak measurement strength $s$ and number of qubits $n$. We observe that the noisy channel under
weak measurement reversal protocol never reduces below the classical resource limit by always generating an average fidelity,
$\langle\cal{F}\rangle_{is} \ge 2/3$. We also observe that the critical damping value leading to the classical limit,
under amplitude-damping, is increased for finite $s$. The plot label UP is for the unprotected state ($s$ = $r$ = 0). }
\label{fig6}
\end{figure}

Under the action of LADC and weak measurement reversal protocol, 
the average fidelity of teleportation using multiparty quantum information splitting ($\langle\cal{F}\rangle_{is}$),
is given by
the expression
\begin{align}
\langle\cal{F}\rangle_{is}=&\frac{1}{3\mathcal{T}}\big[\bar{r}^n + \bar{s}^n(\bar{p} + p\bar{r})^{n-2}(\bar{p}^2 + p^2\bar{r}^2 + p\bar{p}\bar{r})\nonumber\\
 &+(\bar{r}\bar{p}\bar{s})^{n/2}\big].
\end{align}
The average fidelity using information splitting can be maximized with respect to the  reversal weak measurement strength $r$,
keeping the parameters
$p$, $s$, and $n$ fixed.  
Without the protection protocol ($r = 0$ and $s=0$), $\langle\cal{F}\rangle_{is}$ is given by $(2-p\bar{p}+\bar{p}^{n/2})/3$, as obtained in \cite{rafael}.
Figure (\ref{fig6}) is the plot for the numerically optimized average fidelity of teleportation using quantum information splitting. It can be
observed from the
figure that the critical value $p_c^f(n)$ (beyond which the optimized average fidelity is less than or equal to 
the classical teleportation fidelity, $2/3$)
has increased with the initial weak measurement strength $s$, as compared to the average fidelity without any protection from the weak measurement reversal protocol.
It is also observed from Fig. (\ref{fig6}) that the optimal average fidelity always remains above or equal to classical fidelity
that is $\langle\cal{F}\rangle_{is} \ge \cal{F}_{cl}$, for finite weak measurement reversal application.

We observe that both the communication tasks, viz., one sender, many receivers quantum teleportation and multiparty quantum
information splitting, have greater average fidelity under the weak measurement reversal protocol, when the pre-shared resources
suffer from decoherence. The fundamental difference between the two communication tasks is in the optimization to obtain the desired
outcome. While the first protocol requires joint optimization of the encoded states by ($n-1$) Bobs to allow maximal distillation of
the outcome, the second protocol uses multiparty information splitting to decode the state, with the pre-determined Bob performing
the optimization.
%
%
It is argued in \cite{rafael}, that the two protocols utilize different aspects of multipartite quantum correlations to get the average fidelity above the classical upper bound,
using W-like state as a resource. However, in our investigation, we observe that the two quantum communication tasks, using a decohered multi–qubit gGHZ state, do not exhibit any
relation with a specific kind of multiparty entanglement measure.
For example, in the absence of the weak measurement protection protocol, starting with a four qubit decohered gGHZ state, we observe that the MW measure goes to zero
for $p \geq 1/\sqrt{7}$ but fidelity of information splitting is greater than 2/3 for $p \geq 1/\sqrt{7}$.
This clearly shows that the MW measure can not be the only unique resource for the information splitting protocol (c.f. \cite{rafael}).

%
%

\section{Conclusion}
\label{con}

Quantum correlations are the basic resources for various quantum information protocols. From the perspective of futuristic
designs of quantum devices, the scalability of quantum resources is an important aspect of contemporary research. As such,
quantum correlations in multipartite quantum systems need to be harnessed and generated. 
Hence, the study of decoherence protected quantum systems are of fundamental and practical importance from the perspective
of multipartite quantum correlations.

In our study,
we have considered a multipartite, $n$--qubit gGHZ state undergoing local amplitude damping and quantified its mixed state
multipartite quantum correlations using logarithmic negativity and global entanglement. We have then formulated a deocherence
suppression and quantum correlation protection scheme based on weak measurement and reversal technique.
Using analytical characterization and numerical optimization we have evaluated the multipartite quantum correlations under the weak
measurement reversal protocol. We observe that under such a protocol, the multipartite correlations are more robust and do not vanish
for low damping.
In particular we observe an enhanced value of the damping parameter corresponding to the ESD.
To investigate the efficacy
of the weak measurement reversal protocol, we study two quantum communication tasks, namely, quantum teleportation 
in a one sender, many receivers setting
and quantum information splitting through local amplitude damping channel. We observe that the protocol enhances both the 
average fidelity of teleportation and information splitting and prevents 
the channel from performing below the classical upper-limit.

The weak measurement reversal protocol thus strengthens the robustness of global quantum correlations in multipartite
systems under the effect of local amplitude damping. Given the fact that such weak measurements can be experimentally performed,
the protection protocol may prove practically useful in enhancing performance in quantum information tasks,
in presence of noise.

\begin{acknowledgements}

H.S.D. acknowledges University Grants Commission for financial support under the Senior Research Fellowship scheme. H.S.D. thanks 
the Harish-Chandra Research Institute (HRI) for hospitality and support during visits.
The authors thank Aditi Sen(De), Arun Kumar Pati and Ujjwal Sen for various helpful discussions.
U.S. thanks Rafael Chaves and Yong-Su Kim for helpful communications.
 
\end{acknowledgements}


\begin{thebibliography}{999}
\bibitem{comments} This is due to the fact that on tracing out any $m$ sites from the $n$-qubit gGHZ state, the resulting $n-m$ state is always separable.
\end{thebibliography}
\end{document}